\documentclass[preprint,12pt]{elsarticle}

\usepackage{amssymb}
\usepackage{mathrsfs}
\usepackage{diagbox}
\usepackage{placeins}
\usepackage{xcolor}
\usepackage{amsmath}
\usepackage{multirow}
\usepackage{diagbox}
\usepackage{subcaption}
\usepackage{float}
\usepackage{blindtext}
\usepackage{caption}
\usepackage{graphicx}
\usepackage{rotating}

\journal{Applied Acoustics}

\begin{document}

\begin{frontmatter}



\title{An Empirical Wall-Pressure Spectrum Model for Aeroacoustic Predictions Based on Symbolic Regression}


\author[inst1]{Botero-Bolívar, Laura}
\author[inst1]{Huergo, David}

\affiliation[inst1]{organization={ETSIAE-UPM - School of Aeronautics, Universidad Politécnica de Madrid},
            addressline={ Plaza Cardenal Cisneros 3}, 
            city={Madrid},
            postcode={28040}, 
            country={Spain}}

\author[]{dos Santos, Fernanda L. }
\author[inst2]{Venner, Cornelis H.}
\author[]{de Santana, Leandro D. }
\affiliation[inst2]{organization={University of Twente},
            addressline={ Drienerlolaan 5}, 
            city={Enschede},
            postcode={7522 NB}, 
            country={Netherlands}}
\author[inst1]{Ferrer, Esteban}


\begin{abstract}
Fast-turn around methods to predict airfoil trailing-edge noise are crucial for incorporating noise limitations into design optimization loops of several applications. Among these aeroacoustic predictive models, Amiet's theory offers the best balance between accuracy and simplicity. The accuracy of the model relies heavily on precise wall-pressure spectrum predictions, which are often based on single-equation formulations with adjustable parameters. These parameters are calibrated for particular airfoils and flow conditions and consequently tend to fail when applied outside their calibration range. \\
This paper introduces a new wall-pressure spectrum empirical model designed to enhance the robustness and accuracy of current state-of-the-art predictions while widening the range of applicability of the model to different airfoils and flow conditions. The model is developed using AI-based symbolic regression via a genetic-algorithm-based approach, and applied to a dataset of wall-pressure fluctuations measured on NACA 0008 and NACA 63018 airfoils at multiple angles of attack and inflow velocities, covering turbulent boundary layers with both adverse and favorable pressure gradients. Validation against experimental data (outside the training dataset) demonstrates the robustness of the model compared to well-accepted semi-empirical models. Finally, the model is integrated with Amiet's theory to predict the aeroacoustic noise of a full-scale wind turbine, showing good agreement with experimental measurements.
\end{abstract}


\begin{highlights}
\item A new empirical wall-pressure spectrum model based on symbolic regression and experimental data is proposed.
\item The model is validated against experiments of 2D airfoils and compared with other available models.
\item The proposed model shows good accuracy for cases outside the training dataset, and is more robust than other classic models. 
\item The far-field noise of a wind turbine is predicted using the proposed model. The results compare well with field measurements. 
\end{highlights}


\end{frontmatter}


\section{Introduction}\label{sec:intro}
%
Rotating blades found in wind turbines or propellers generate noise that can cause health and physiological problems in humans in close proximity. More generally, anthropogenic noise can have a non-negligible impact in animals, disrupting essential activities such as communication, reproduction, or predator avoidance~\cite{Abrahamsen2012,KURT2016, Gotz2009, erbe2022}. Therefore, noise regulations have arisen in several fields~\cite{Regulation, icao2016, Baudin2015,DAVY2018} to limit noise generation and reduce the impact on the environment. Modeling and predicting the noise produced by an airfoil with fast turn-around methods is crucial to incorporate noise limits in the design phase of new rotating blades.

Flow-induced noise is the most important noise source in several applications such as wind turbines, aircraft, and ships. Trailing-edge noise, also known as airfoil self-noise, is the minimum amount of noise that an aerodynamic surface can produce~\cite{roger2005}. It is caused by the interaction of the turbulent boundary layer with the airfoil finite trailing edge. The convecting turbulence within the boundary layer causes wall-pressure fluctuations. Close to the trailing edge, these fluctuations are scattered to the far field as noise due to the sudden change of impedance at the trailing-edge discontinuity. Indeed, there is a high correlation between the wall pressure fluctuations spectrum (WPS) close to the trailing edge and the far-field noise spectrum. Consequently, semianalytical noise prediction methods are often based on the wall-pressure spectra. This is the case of  Amiet's theory~\cite{amiet1976TE}. Measurements of wall pressure fluctuations close to the trailing edge during experiments is extremely challenging due to the thin trailing edge thickness of wind tunnel models. Numerical simulations are also challenging when trying to resolve turbulent structures near the airfoil walls, resulting in expensive simulations that cannot be used during design optimization procedures. Therefore, finding an accurate and general model for the turbulent boundary layer WPS is fundamental for the accurate and rapid assessment of trailing-edge noise.  

The WPS beneath a turbulent boundary layer has a broadband nature and contains information on the flow structures across the boundary layer of both the inner and the outer regions~\cite{Devenport}. The low-frequency range is caused by the large structures present in the outer part of the boundary layer and therefore scales with the boundary layer thickness ($\delta$) and the edge velocity ($U_e$)~\cite{Devenport}. The smaller turbulent structures, located close to the wall, are responsible for the wall-pressure fluctuations in the high-frequency range, which scale with the viscous length defined as $\nu/u_\tau$, where $\nu$ is the kinematic viscosity and $u_\tau$ denotes the friction velocity~\cite{Devenport, goody2004}. The mid-frequency range is generated by both types of structures coexisting in the boundary layer, and is known as the overlap frequency range. Based on these scalings, several semi-empirical models have been proposed for the turbulent boundary layer WPS~\cite{goody2004,chase1980,rozenberg2012, catlett2016, kamruzzaman2015semi, hu2016, parchen1998progress}. 

Initially, \citet{Corcos1964_model}, \citet{goody2004}, and \citet{chase1980} proposed semi-analytical models using empirical constants based on measurements of a flat plate under zero pressure gradient (ZPG). \citet{catlett2016} extended Goody's model by testing three different trailing-edge configurations that change the stream-wise pressure gradient. \citet{rozenberg2012} proposed an empirical model based on Goody's model considering ZPG and adverse pressure gradients (APG). For both cases, the pressure gradient is considered in the form of Clauser's equilibrium parameter. 
The main drawback of Rozenberg's model is that it relies on the velocity fluctuations across the boundary layer, which are difficult to obtain with fast turn-around methods, e.g., XFOIL.
Based on Rozenberg's model, \citet{kamruzzaman2015semi} proposed a model considering both favorable pressure gradient (FPG) and APG based on experimental measurements of several airfoils at different angles of attack. \citet{LEE20211} proposed a model combining Rozenberg's, and Kamruzzamn's models by incorporating the pressure gradient to better define the transition from mid to high frequency in the WPS. A significant difference between Kamruzamann and Lee's model to Rozenberg's is that the former two use the wall shear stress ($\tau_w$) instead of the maximum across the boundary layer ($\tau_\mathrm{max}$) used in Rozenberg's model. \citet{hu2016} proposed their model based on measurements of a NACA~0012. They account for pressure gradient with the boundary layer shape factor instead of Clauser's equilibrium parameter. Recently, \citet{moreau_GWPS} proposed a generalized model based on Goody's model considering direct numerical simulations (DNS), large eddy simulations (LES), and experiments for attached, separated, and reattached boundary layers. The characteristic of this generalized model is the use of velocity fluctuations across the boundary layer (similar to Rozenberg's model), which are difficult to model using fast engineering tools. \citet{Dominique_WPS_ANN} and \citet{arroyo2024_ANN_WPS} used artificial neural networks as an alternative approach to improve the robustness of the WPS model and better predict cases with strong pressure gradients. The neural network was used to predict boundary layer velocity profiles to later model the WPS. All the models mentioned so far are presented as a single equation with eight parameters that give the shape of the spectrum besides scaling factors for the level and the frequency. Therefore, the contribution of each semiempirical model has focused on modeling the eight parameters and the scaling factors instead of proposing a completely new model.%

Despite the significant effort of the aeroacoustic community, there is no consensus on a WPS model that is general and usable outside its calibration range and can be used with confidence to predict trailing-edge noise to any case~\cite{kuccukosman2018, kissner2022}. The main drawback of the empirical models available in the literature is the strong dependence on the data used to tune the aforementioned eight parameters of the model, limiting their suitability for actual scenarios, such as in rotating wind turbine blades. The main limitation is related to the use of one unique equation to model the WPS for the entire frequency range, which requires encapsulating a variety of frequency-dependent physics in only one equation. As mentioned before, each frequency range of a turbulent boundary layer WPS is generated by different flow structures that scale with different boundary layer quantities. When using a single equation to model the entire frequency range of the WPS, the physical principle of the generation of wall pressure fluctuations is oversimplified. An alternative approach was followed by \citet{fritsch2023modeling}, who used a symbolic regression algorithm to define functions that are coupled in a single equation. Each function has a relative importance within the frequency range in an attempt to better model the different decays present in a WPS. In what follows, we will consider a similar approach.

To overcome the limitations of the available (semi-) empirical models, we propose an empirical model for the wall-pressure spectrum that: 
\begin{itemize}
    \item Predicts independently curves for three frequency bands, which are blended together into one model. This approach enables capturing different physics and renders the model more robust and generalizable (e.g., making the model more robust to predict the curvature around the inflection point in the low-frequency range); 
    \item the model is generated using an AI-based symbolic regression, which is a machine learning method capable of unveiling physically interpretable analytical equations from data, and reduces the need to incorporate prior knowledge about the physical phenomenon (i.e., we do not need to provide any specific equation);
    \item the considered database consists of experimental measurements that include two different types of airfoil geometries, Reynolds numbers and FPG and APG, which make the model robust to provide a better prediction for cases outside the training dataset.
\end{itemize}

Additionally, we force the model to use as input simple boundary layer parameters, i.e., boundary layer displacement thickness ($\delta^*$), boundary layer momentum thickness ($\theta$), and friction coefficient ($c_f$), which are parameters that can be obtained with fast turn-around methods such as XFOIL~\cite{drela1989}, without the need of running costly numerical simulations. 


This paper is organized as follows. Section~\ref{Sec:meth} addresses the methodology followed to propose the WPS model, including the definition of the model, the machine learning approach, and wind tunnel experiments. Section~\ref{sec:results} presents the WPS model. Section~\ref{sec:val} shows the validation of the model with other airfoil data available in the literature and a complete wind turbine. Finally, Section~\ref{sec:con} presents the main conclusions of this research. 

\section{Methodology}\label{Sec:meth}

\subsection{Wall-pressure spectrum definition}
\begin{figure}[h]
\centering
\includegraphics[width=0.6\textwidth]{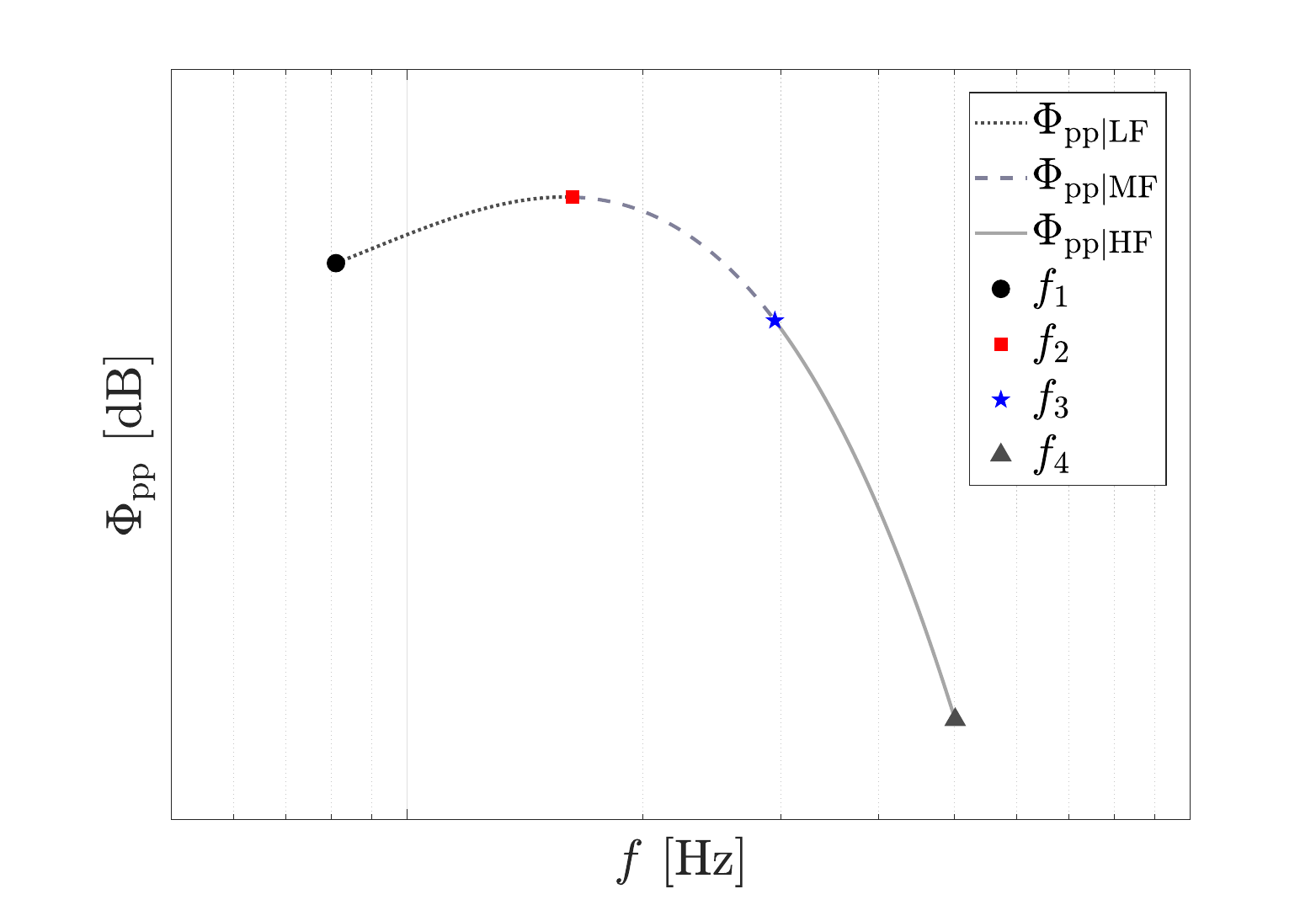}
\caption{Frequency bands of a generic WPS.}\label{fig:spect_def}
\end{figure}

The WPS ($\Phi_\mathrm{pp}$) is divided into three frequency bands: low-frequency, mid-frequency, and high-frequency, that is, $\Phi_\mathrm{pp|LF}$, $\Phi_\mathrm{pp|MF}$, and $\Phi_\mathrm{pp|HF}$, respectively, which are delimited by $f_2$ and $f_3$, as shown in Fig.~\ref{fig:spect_def}. $f_2$ limits the low- and mid-frequency bands, and is defined as the frequency of the maximum spectral level along the entire frequency range. $f_3$ limits the mid- and high-frequency ranges. $f_3$ is defined as the frequency where the decay of the WPS level as a function of frequency changes significantly, suggesting that a different physical phenomenon dominates the WPS.

To obtain $f_3$, the experimental spectrum was divided into frequency bins of 500~Hz; the center of the bin was progressively moved by 240~Hz. The spectrum slope for each bin was calculated by fitting a first-order polynomial to the logarithmic values of the spectral level and the frequency. $f_3$ is considered where the slope deviates by more than 5\% from the value of the previous bin. $f_1$ and $f_4$ define the frequency range in which the WPS is calculated. Those frequencies do not affect the model itself but are needed to blend the curves of each frequency bands into a single curve. $f_1$ is defined as $0.5f_2$ and $f_4$ is defined as the highest frequency limit where the WPS is calculate. For the training dataset, it is defined as 5~kHz to avoid contamination of electronic noise during the experimental measurements. 

Each frequency band was modeled independently on the logarithmic scale using the symbolic regression approach, where we obtained three equations. The inputs given to the symbolic regression algorithm to model the spectrum are the inflow velocity ($U$), boundary layer thickness ($\delta$), boundary layer displacement thickness ($\delta^*$), boundary layer momentum thickness ($\theta$), wall shear stress ($\tau_w$), and the pressure gradient ($\mathrm{d} C_p/\mathrm{d} x$). 
This set of parameters was chosen for two main reasons. The first one is that all these variables can be easily obtained for a specific airfoil without running expensive simulations (e.g., using fast panel method solvers such as XFOIL). Secondly, these parameters are chosen to be as general and independent as possible. Although other models include additional variables (such as $\Delta = \delta/\delta^*$, or $\beta = (\delta^*/\tau_w)(\mathrm{d}p/\mathrm{d}x)$), they are not retained in this work since they can be defined as a combination of the previous selected ones and therefore could introduce a bias in the model, providing a solution which is not optimal.

The limiting frequencies ($f_2$ and $f_3$) are modeled using the same inputs, and the equations obtained for each frequency band are combined through a weighted average as a function the frequency, as will be explained in the following sections. 

\subsection{Symbolic regression approach}
Symbolic regression (SR) has emerged in recent years as an alternative to traditional regression algorithms to derive models from data (fitting data). While the traditional approach is based on a user-defined function, with some tunable coefficients that can be adjusted to minimize the error between the function and the data, the SR approach allows us to discover new functions from scratch. The main idea is to provide mathematical operators to the algorithm that can be combined automatically to define the best equation to fit the data. This way of solving problems involves a change in the current paradigm and provides an alternative to rediscover formulations of physical laws~\cite{tenachi2023deep}, among other applications. A main advantage of SR is that it does not require a priori specification of the model, and hence it is not affected by human bias or unknown gaps in domain knowledge.
Although this family of algorithms was developed in the early 1990s, recent advancements in machine learning and the rapid improvement of computing power have fostered the use of SR for different fields; such as wind turbine wake modeling~\cite{wang2024discovering}, astrophysics~\cite{tenachi2023deep}, turbulence~\cite{vaddireddy2020feature}, discovery of missing terms in differential equations~\cite{lozano2019towards}, material science~\cite{wang2019symbolic} and prediction of solar power production~\cite{quade2016prediction}, among others. The application of SR in the field of acoustics is scarce, but some advances have been made in the prediction of sound propagation in shallow water
environments~\cite{tohme2024isr} and in the modeling of noise generation for porous airfoils~\cite{sarradj2014symbolic}. Although some attempts have been made in the reconstruction of the WPS model through symbolic regression~\cite{dominique2021inferring, fritsch2023modeling} and neural networks (NN)~\cite{Dominique_WPS_ANN}, these models are difficult to generalize (for SR-based approaches) and lack interpretability (in the case of using NN).

In this work, SR is used to discover mathematical expressions to model the turbulent boundary layer WPS of two airfoils. The algorithm training is performed on the basis of experimental data (detailed in the following sections). Furthermore, the set of allowed variables are inflow conditions and boundary layer parameters that can be easily defined/obtained, allowing for a generalizable approach. 

Symbolic regression algorithms are based on the main idea of continuously constructing new analytical functions, constrained by the user's input, to fit a set of data. These functions are created by using a set of operators (e.g., $+$, $-$, $*$, $/$,...), functions (e.g., $\mathrm{abs}()$, $\sin()$, $\log()$,...) and variables (e.g., $x$, $t$,...) that the user must decide beforehand. The resulting solution is formed by a combination of these operators and variables in a specific order. 
Another key aspect of SR is the \emph{complexity} of the solution. If no constraints were applied to limit the complexity, very long equations with poor interpretability could be created. Therefore, the complexity of the final solution is usually constrained by defining the maximum length of the new functions. Furthermore, low-complexity functions are less prone to overfitting, leading to more generalizable solutions.

During training, new functions are created continuously, guided by an optimization algorithm, with the final objective of minimizing the loss between the discovered solution and the real data.
The generation process consists of solving an optimization problem, which is usually handled by genetic algorithms, where the function itself is mutated by modifying its components and the interrelation among them. In the present work, we apply the methodology implemented by \citet{cranmer2023interpretable}, who provides an open-source code with a multi-population evolutionary algorithm to solve the SR problem in an efficient and modular way. This approach has been thoroughly tested and is capable of exceeding several state-of-the-art open-source symbolic regression algorithms. This original implementation is based on genetic algorithms for optimization~\cite{brindle1980genetic} at its core, but includes additional features to improve performance, such as simulated annealing~\cite{kirkpatrick1983optimization}, classic Broyden–Fletcher–Goldfarb–Shanno (BFGS) optimization algorithms~\cite{broyden1970convergence}, and a novel adaptive parsimony metric. 
These improvements are combined with an asynchronous programming approach and multithreading parallelization, enhancing the overall performance and reducing the simulation time.

\subsection{Experimental database}\label{sec:dataset}
\begin{figure}[htb!]
    \centering
    \begin{minipage}{0.48\textwidth}
    \centering
    \includegraphics[width=0.9\textwidth]{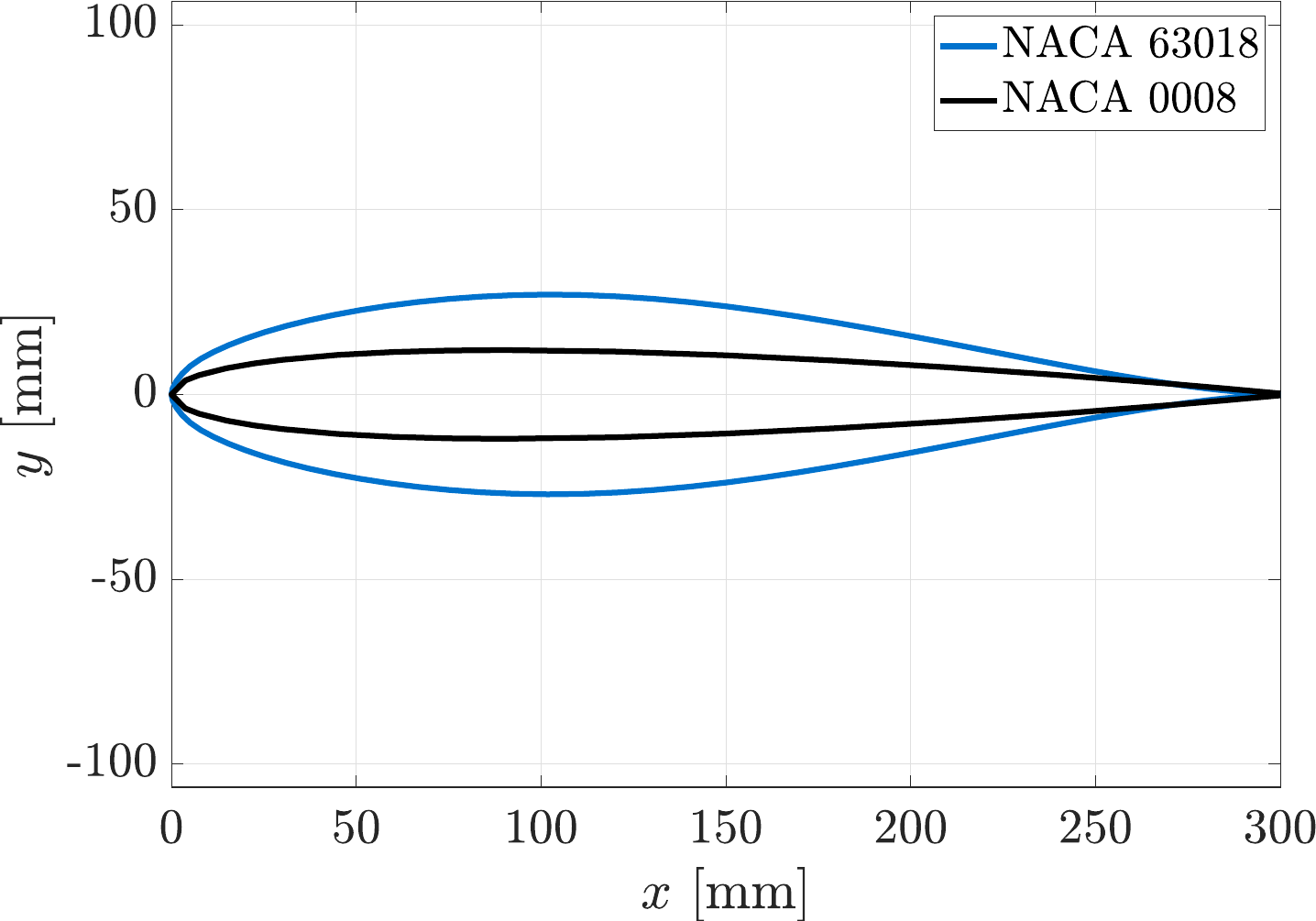}
    \caption{\label{Fig: airfoils} Airfoils' coordinates.} 
    \end{minipage}
	\hfill
	 \begin{minipage}{0.48\textwidth}
        \centering
        \includegraphics[width=0.9\textwidth]{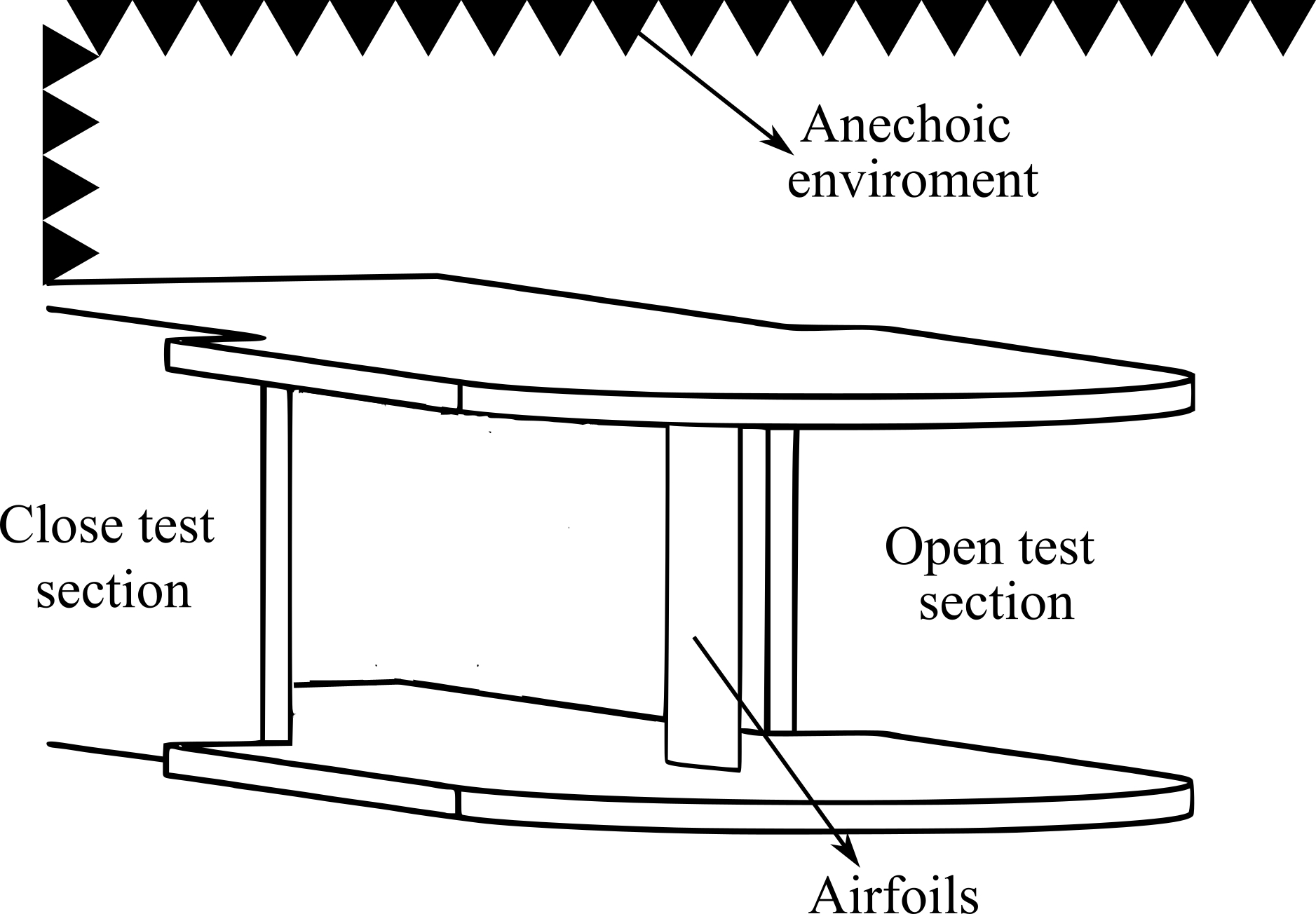}
        \caption{\label{Fig: setup} Schematic view of the experimental set-up.} 
    \end{minipage}
\end{figure}

The symbolic regression algorithm is applied to wall-pressure fluctuation measurements of a NACA~0008 and a NACA~63018 of 300~mm chord ($c$). The geometry of the airfoils is shown in Fig.~\ref{Fig: airfoils}. The WPS is measured at 97\% of the chord of each airfoil on both the suction and pressure sides. The test conditions for each airfoil are shown in Table~\ref{tab:conditions}, where $\alpha_g$ is the geometric angle of attack (the effective angle of attack $\alpha_e$ is calculated as explained in Section~\ref{sec:corrections}), $U$ is the inflow velocity, and $Re$ is the Reynolds number based on $U$ and $c$. These conditions provide a total of 76 cases. Specifically, for the high-frequency band, only the cases for which there were enough data were used to train the model, i.e., cases where $f_3<$4~kHz. The total number of cases for this frequency range is 47. 
\begin{table}[h!]
    \centering
    \begin{tabular}{|c|c|c|c|}
    \hline
        Airfoil & $\alpha_g$ [\textdegree] & $U$ [m/s]  & $Re$ [-]\\
        \hline
      \multirow{3}{*}{NACA~0008 }  & 0:0.5:6   & 30 & 7.3 $\times$ 10\textsuperscript{5} \\
        & 0, 3, 5  & 10, 30 & \{2.4, 7.3\} $\times$ 10\textsuperscript{5} \\
       & 0  & 10:5:45  & \{2.4:1.2:10.1\} $\times$ 10\textsuperscript{5}  \\
       \hline
       \multirow{2}{*}{NACA~63018} & 0:1:7   & 10, 30  &\{2.4, 7.3\}$\times$ 10\textsuperscript{5} \\
       & 0  & 10:5:45& \{2.4:1.2:10.1\} $\times$ 10\textsuperscript{5}\\
       \hline
    \end{tabular}
    \caption{\label{tab:conditions}Experimental test conditions.}
\end{table}

\subsection{Wind tunnel experiments}
\subsubsection{Wind tunnel}
The experiments were carried out in the aeroacoustics wind tunnel of the University of Twente. The facility is a closed circuit wind tunnel, which has a 0.9~m width and 0.7~m height close and open test sections. The wind tunnel has a contraction ratio of 10:1 and seven screens that keep the turbulence intensity below 0.08\% with velocities up to 60~m/s~\cite{deSantana2018}. The test section is enclosed by an anechoic chamber of 6~m $\times$ 6~m $\times$ 4~m. The empty anechoic chamber has a cut-off frequency of 160~Hz. The temperature in the test section is controlled at approximately 20\textdegree C. The airfoils were vertically installed in the open test section using two rotatory end plates, allowing changes in the angle of attack of 0.5\textdegree~precision. More information on the wind tunnel can be found in~\citet{deSantana2018}. Figure~\ref{Fig: setup} shows a schematic representation of the experimental setup.

\subsubsection{Wind tunnel corrections}\label{sec:corrections}
 The wind tunnel experiments were conducted in an open jet test section. Therefore, the geometric angle of attack ($\alpha_g$) was corrected to obtain the effective angle of attack ($\alpha_e$). Corrections were based on the approach proposed by \citet{brooks1989}.
 \begin{equation}\label{Eq:WT_cor}
     \sigma = (\pi^2/48)(c/W)^2; \qquad d\sigma = (1 + 2\sigma)^2 + \sqrt{12\sigma}; \qquad \alpha_e = \alpha/d\sigma;
 \end{equation}
 where $W$ is the wind tunnel width. All data and analyses are presented in relation to the effective angle of attack.

\subsubsection{Boundary layer transition}
 The boundary layer transition was forced at 6.5\% of the chord, using zigzag strips of 60\textdegree~zigzag angle and 12~mm width. The trip height ($k$) was varied with the inflow velocity and the angle of attack to keep the ratio $k/\delta_k$ between 0.4 and 0.6 for all cases, where $\delta_k$ is the boundary layer thickness in the trip location. $\delta_k$ was estimated using XFOIL simulations~\cite{drela1989}, using the approach explained in Section~\ref{sec:xfoil_sim}. For this particular case (to calculate $\delta_k$), XFOIL simulations were conducted using natural transition with the critical number, calculated based on the inflow turbulence (Tu~=~0.08\%):
\begin{equation}\label{Eq:Ncrit}
    N_{\mathrm{crit}}=-8.43-2.4\log(Tu/100). 
\end{equation}

\subsubsection{Measurement of wall-pressure fluctuations}
The airfoils are instrumented with 82 remote microphone probes (RMP) distributed along the chord and the span. The wall-pressure fluctuations are measured on both the suction and pressure sides at 95\% of the chord. Twelve microphones are distributed along the span at 97\% of the chord. The statistical similarity of the measurements of the twelve microphones was compared. The WPS matched within 2~dB for all the microphones.

Remote microphone probes consist of pinholes of 0.3~mm diameter on the airfoil surface connected to a unique stainless steel tube of 1.6~mm inner diameter. A tube assembly outside the model connects the tubes to the microphones and the anechoic termination, i.e., a plastic tube of 1.6~mm inner diameter and 3~m long. The extreme of the anechoic termination was sealed to avoid airflow passing through the tube because of the difference in pressure between the airfoil surface and the environment. On the side of the tube assembly, Knowles FG~23329-P07 microphones are connected. The junctions of the stainless steel tube, the anechoic termination, and the microphones with the tube assembly were sealed to prevent leakage, which was verified during the calibration of each RMP. 
The calibration of RMPs is explained in~\ref{sec:rmp_cal}. More information on the technique can be found in \citet{dossantos_rmp}.

Four National Instruments PXIe-4499 Sound and Vibration modules installed on a NI PXIe-1073 chassis were used to sample the surface microphones. The microphone data was acquired during 30~s with a sampling frequency of 65536 (2\textsuperscript{16})~Hz. The WPS was calculated by adopting the Welch method, using a window size of 2\textsuperscript{14} samples (7.5~s) and a Hanning windowing method with 50\% overlap, which resulted in a bin size of 8~Hz. The WPS is presented in dB, calculated according to \citet{Devenport}. The reference pressure and delta frequency for normalizing the power spectral density were 20 \textmu Pa and 1~Hz.

\subsection{XFOIL simulations}\label{sec:xfoil_sim}
XFOIL simulations were performed replicating the conditions of the wall-pressure fluctuations measurements to obtain the parameters of the boundary layer close to the trailing edge needed to propose the model. The input parameters for XFOIL were the airfoil geometry, the Reynolds number, calculated on the basis of the airfoil chord and inflow velocity, the Mach number, the effective angle of attack, and the location of the forced transition (i.e., 0.065). The boundary layer parameters were extracted from XFOIL simulations at $x/c=$0.95, i.e., the same location of the wall-pressure fluctuations measurements. 

XFOIL calculates the displacement thickness ($\delta^*$), momentum thickness ($\theta$), and skin friction coefficient ($C_f$). The boundary layer thickness ($\delta$) is calculated as~\cite{drela1989}:
\begin{equation}\label{eq:delta_xfoil}
    \delta~=~\theta \left( 3.15 + \frac{1.72}{H_k-1}+\delta^* \right),
\end{equation}
where $H_k$ is the kinematic shape parameter:
\begin{equation}\label{eq:kinematic_shape_parameter}
    H_k~=~\frac{H-0.290 M_e^2}{1+0.113 M_e^2},
\end{equation}
with $M_e$ the boundary layer edge Mach number, and $H$ the shape parameter given as $H=\delta^*/\theta$. For low Mach number flows, $H_k$ given by Eq.~\ref{eq:kinematic_shape_parameter} reduces to $H$. This approximation is used in this study.

The friction velocity, $u\tau$, is calculated as:
\begin{equation}\label{Eq:u_tau}
u_\tau = \sqrt{\tau_w/\rho},
\end{equation}
where $\tau_w= 0.5\rho U^2 C_f$.
Finally, the pressure gradient $\mathrm{d}C_p/\mathrm{d}x$ is calculated using the central differencing scheme, with the $x/c=$0.95 as the central value. 

XFOIL simulations are validated against experiments by comparing $\delta$, $\delta^*$, $\theta$, and $u_\tau$ close to the trailing edge, and $C_p$ distribution along the chord. This validation is shown in~\ref{sec:app_Xfoil_val}. There is a good agreement between XFOIL and the experiments.  

\section{Wall-pressure spectrum model}\label{sec:results}
The SR-based WPS model is defined based on a set of operators, functions, and variables that the SR algorithm is allowed to use when generating new functions. In this case, we will try to minimize the number of available degrees of freedom, as the objective is to obtain a simple and interpretable final solution:
\begin{enumerate}
    \item \textbf{Operators:} $+$, $-$, $*$ and $/$.
    \item \textbf{Functions:} $\mathrm{pow2}(x)=x^2$, $\mathrm{abs}(x)=|x|$.
    \item \textbf{Variables:} $f$, $U$, $\delta$, $\delta ^*$, $\theta$, $\tau _w$ and $\mathrm{d}C_p/\mathrm{d}x.$ 
\end{enumerate}
Furthermore, the maximum complexity of an equation is set to 25, where the use of each operator, function, or variable increases the complexity by 1 unit. Also, the algorithm is allowed to include real constants, but each constant will increase the complexity by 2 units.

The initial population of functions is formed by 200 individuals and the multi-population genetic algorithm runs 50 iterations for each function to be discovered. The loss during training is computed as the $L_2$ norm $|| x - x^* ||_2$ between the original data $x$ and the generated function $x^*$, within the current frequency band. %

The WPS in each frequency band is modeled in logarithmic scale and consequently expressions must begin with a $10 \log_{10}(\bullet)$. Moreover, on the basis of the experimental data, the high-frequency region has an almost linear decay on a logarithmic scale; hence, we force a linear decay in the high-frequency band. 

As a result of the SR process, the following set of equations are discovered to model/calculate the WPS in each frequency band (see Fig.~\ref{fig:spect_def}), and the limit frequencies: 
\begin{equation} \label{eq:model}
\begin{split}
f_1 & = 0.5f_2; \\
f_2 & =  55.67 + 0.045~U/\delta^*; \\
f_3 & = 948.4 + 0.26~U/\delta; \\
f_4 & = 5000; \\
\Phi_\mathrm{pp|LF}  &= 10 \log_{10} \left( 0.16 \left( \delta U \right)^2 \right); \\
\Phi_\mathrm{pp|MF}(f) &= 10 \log_{10} \left( \left( \frac{f}{f_2} \right)^{\frac{b}{3}} \right) + \left( \frac{\Phi_\mathrm{pp|LF} + offset_\mathrm{MF}}{2} \right); \\
b &= -\left| \tau_w- 1804.5\theta \right|; \\
offset_\mathrm{MF} &= 10 \log_{10} \left( 4.41 \cdot 10^{-4} \frac{ \left( U  \left( 6.78 \cdot 10^{-4}+ (\tau_w - \delta)^2 \right) \right)}{f_2} \right);\\
\Phi_\mathrm{pp|HF}(f) &= 10 \log_{10} \left( \left( \frac{f}{f_3} \right)^{b} \right) + \Phi_\mathrm{pp|MF}(f_3).
\end{split}
\end{equation}

The frequency $f_2$ depends on the inflow velocity and boundary layer displacement thickness, whereas $f_3$ depends on the boundary layer thickness instead. The equation for the low-frequency range ($\Phi_\mathrm{pp|LF}$) is a constant (non-dependent on the frequency, $f$) that gives the level of the WPS, as a function of the boundary layer thickness and the inflow velocity. The high-frequency band ($\Phi_\mathrm{pp|HF}$) is a logarithmic straight line with a decay $b$ that depends on the wall shear stress and momentum thickness. The decay for the mid-frequency band ($\Phi_\mathrm{pp|MF}$) is one-third of that of the high-frequency band. Both equations have an offset to set the level of the curve. The offset in the mid-frequency range ($offset_\mathrm{MF}$) depends on velocity, wall shear stress, boundary layer thickness, and $f_2$. To reduce the discontinuity of the curve at $f_2$, where the LF and MF regions blend together, the mid-frequency offset is averaged with the value of the low-frequency band. The offset in the high-frequency band is simply the value of the WPS given by the expression of the mid-frequency range evaluated at $f=f_3$ ($\Phi_\mathrm{pp|MF}(f_3)$).

\begin{figure}[H]
	\centering
	\begin{subfigure}[c]{0.45\textwidth}
		\centering
		\includegraphics[width=\textwidth]{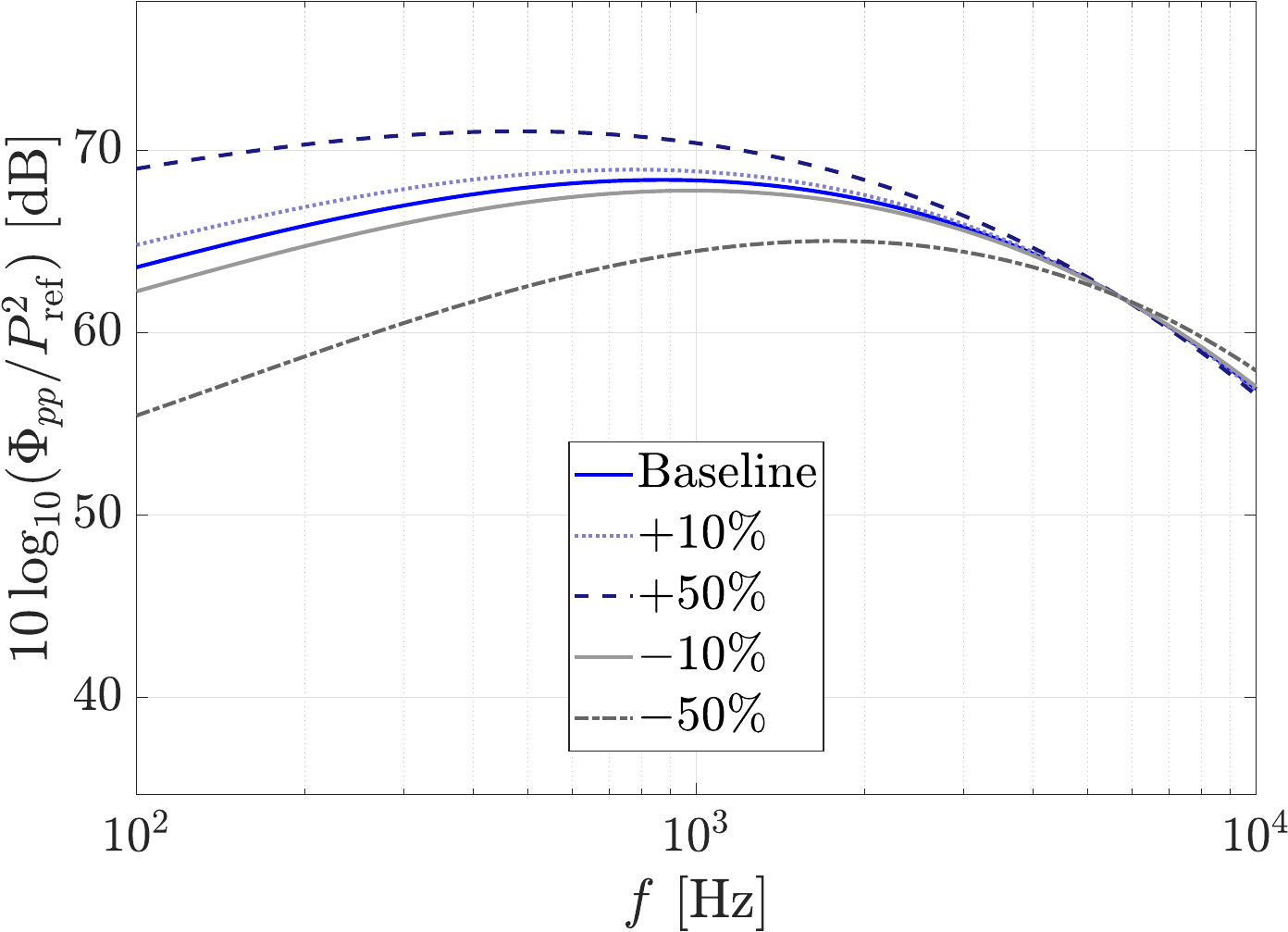}
            \caption{Boundary layer thickness, $\delta$}
		\label{subfig:sens_delta}
	\end{subfigure}
	\hfill
	\begin{subfigure}[c]{0.45\textwidth}
		\centering
		\includegraphics[width=\textwidth]{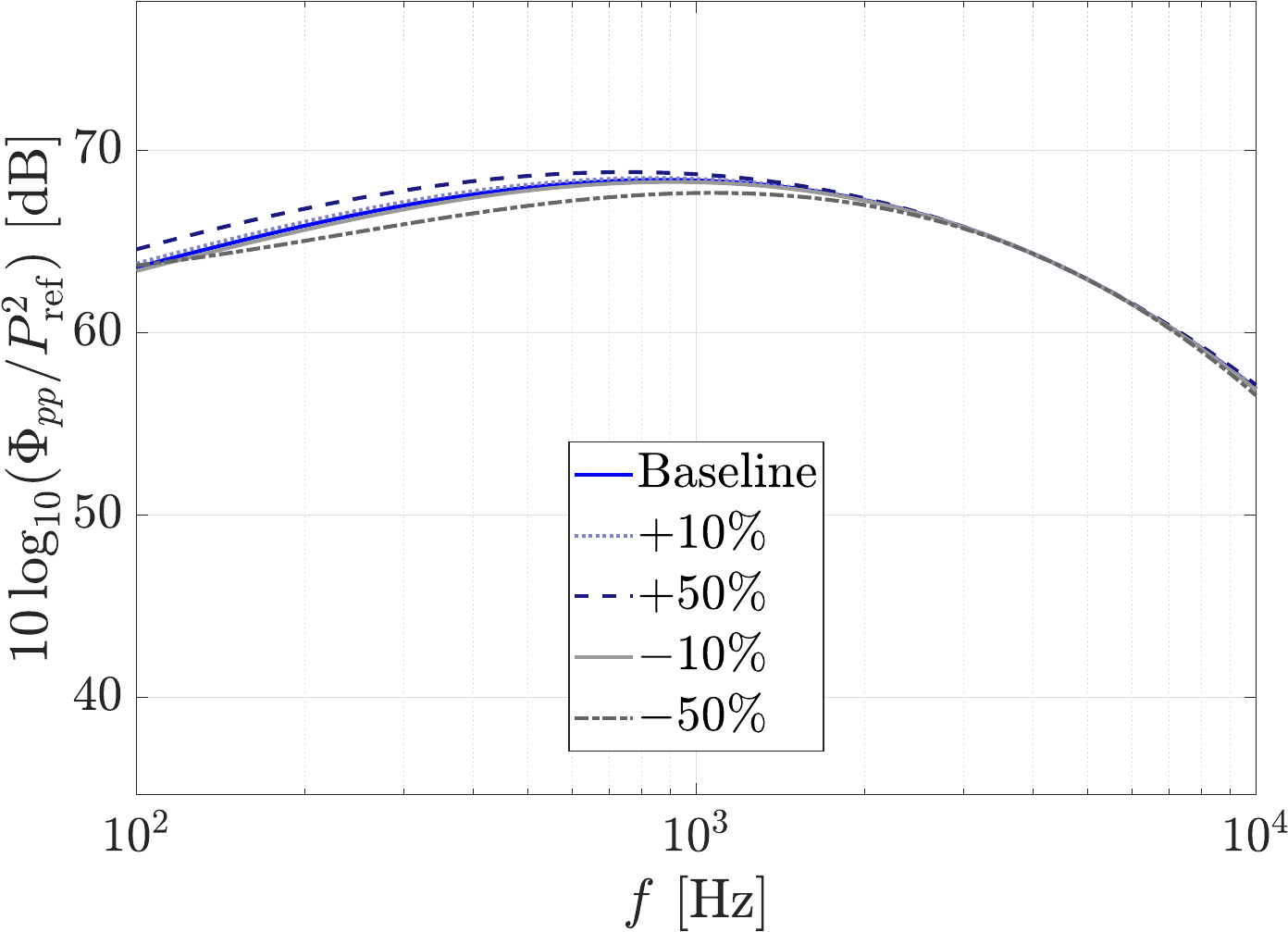}
		 \caption{Boundary layer displacement thickness, $\delta^*$}
		\label{subfig:sens_delta_s}
	\end{subfigure}	
    \vspace{5mm}
	\begin{subfigure}[c]{0.45\textwidth}
		\centering
		\includegraphics[width=\textwidth]{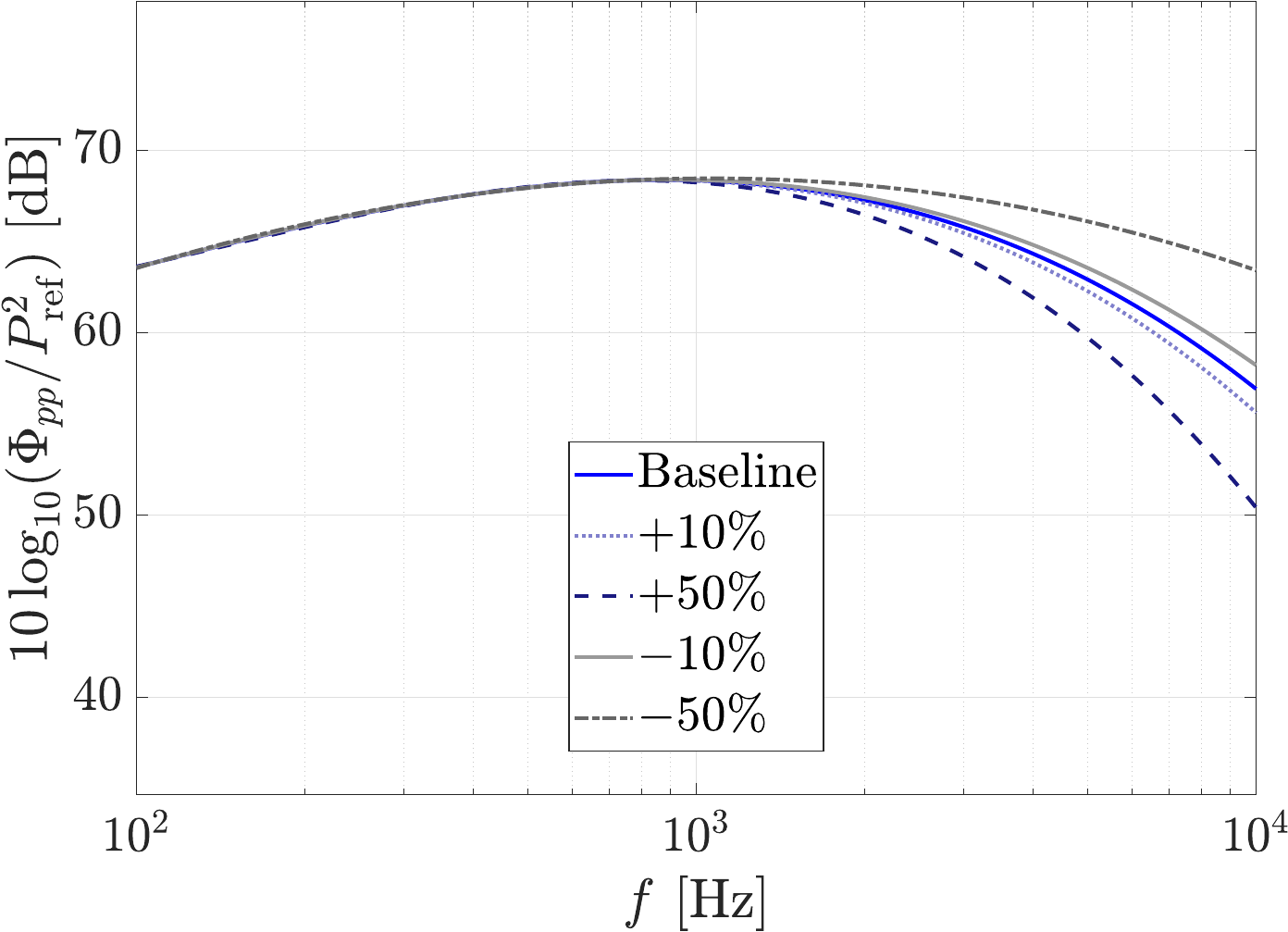}
        \caption{Boundary layer momentum thickness, $\theta$}
		\label{subfig:sens_theta}
	\end{subfigure}
	\hfill
	\begin{subfigure}[c]{0.45\textwidth}
		\centering
		\includegraphics[width=\textwidth]{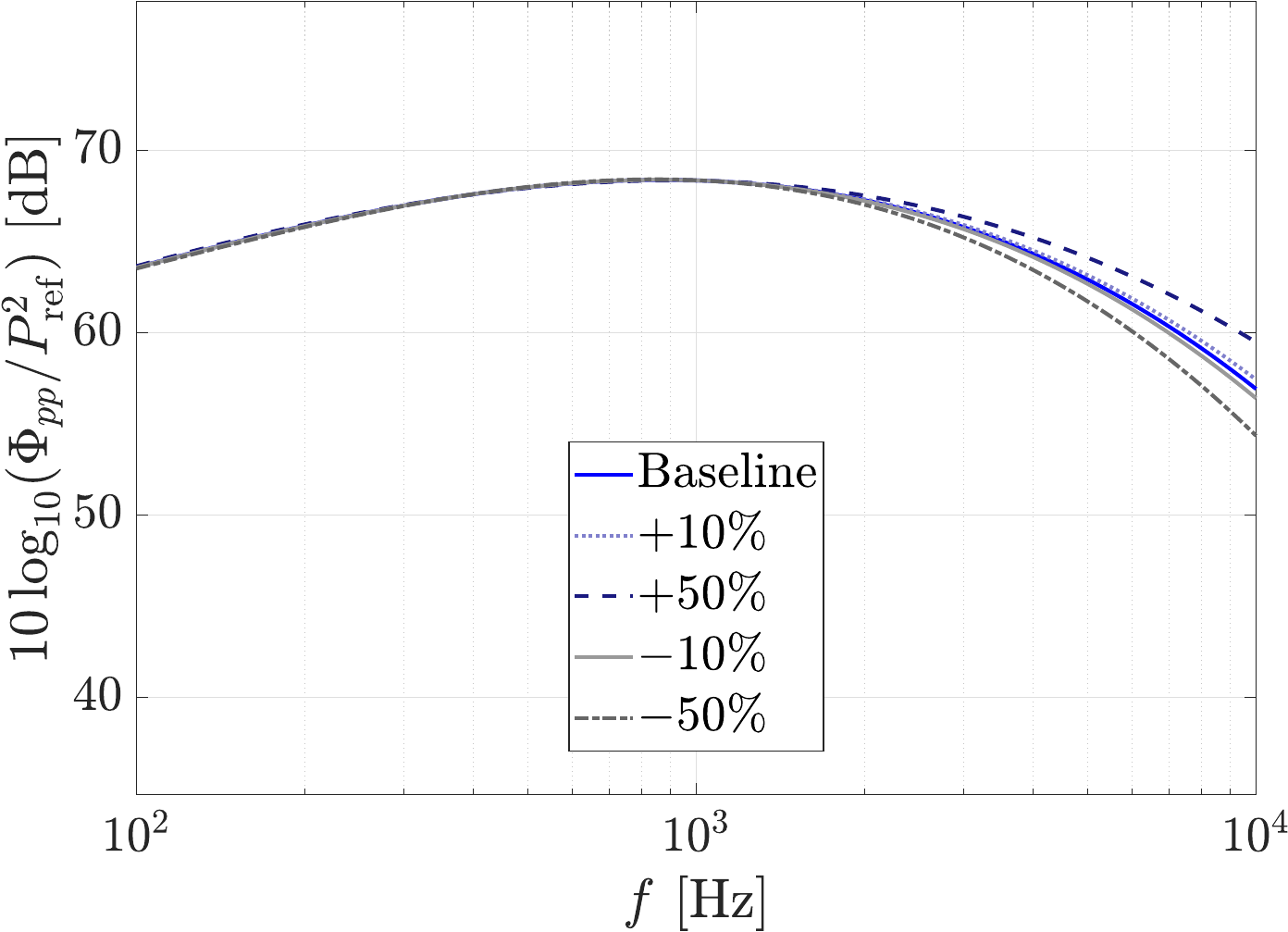}
		\caption{Wall shear stress, $\tau_w$}
		\label{subfig:sens_tau_w}
	\end{subfigure}
         \vspace{5mm}
	\begin{subfigure}[c]{\textwidth}
		\centering
		\includegraphics[width=0.45\textwidth]{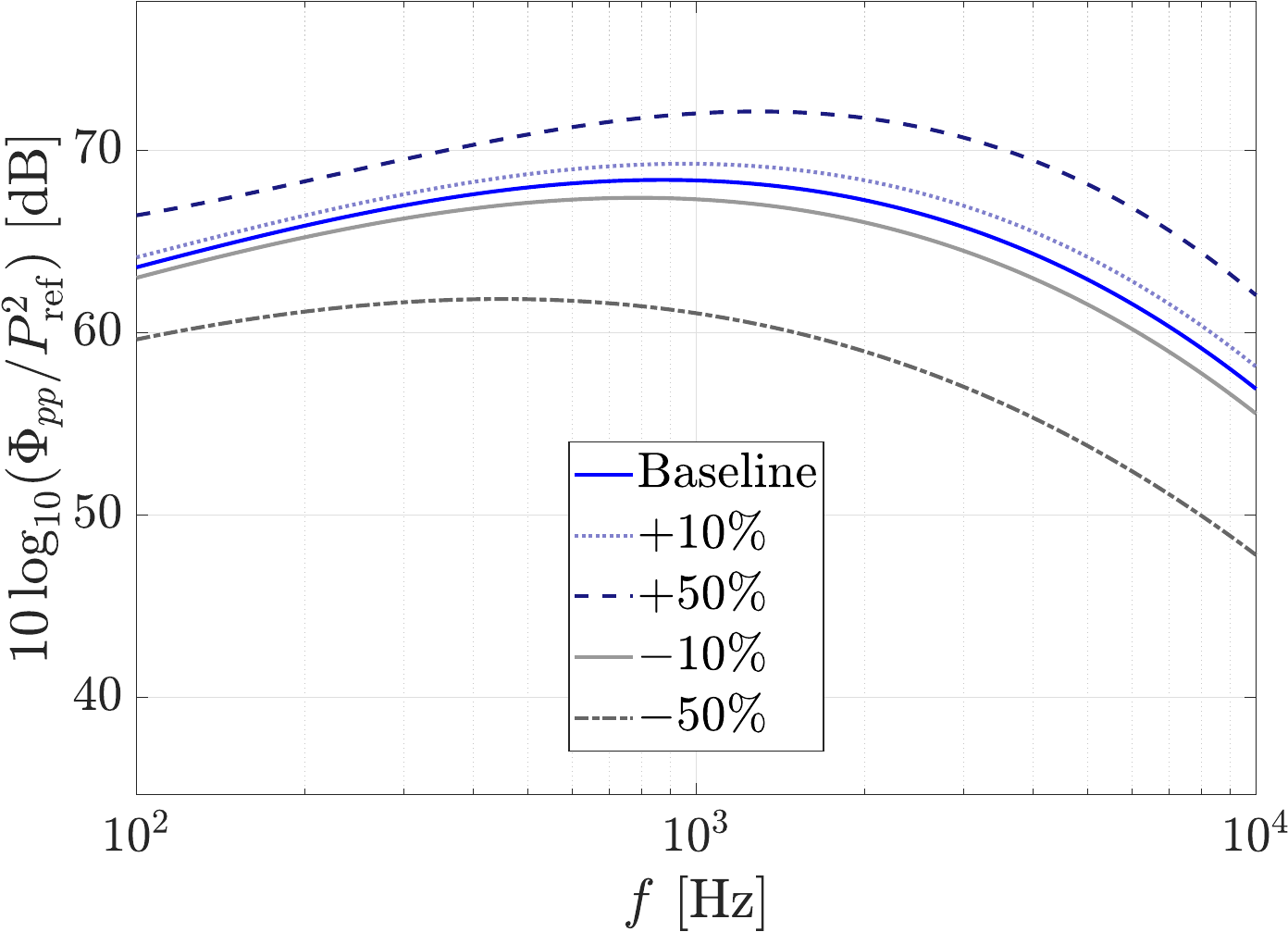}
            \caption{Inflow velocity, $U$}
		\label{subfig:sens_u}
	\end{subfigure}
	\caption{Sensitivity of the boundary layer parameters on the predicted wall-pressure spectra.}
	\label{fig:sens_anal}
\end{figure}

The model obtained is shown in Fig.~\ref{fig:sens_anal}, where the influence of the boundary layer parameters on the WPS is detailed. The boundary layer thickness ($\delta$) and displacement thickness ($\delta^*$) affect the level and position of the maximum spectral level, while there is no effect in the high-frequency band. As the boundary layer thickens, the energy content of the WPS increases and shifts towards lower frequencies. In contrast, momentum thickness ($\theta$) and wall-shear stress ($\tau_w$) affect the level and decay of the high-frequency range. The energy content of the WPS increases and shifts towards higher frequencies as $\tau_w$ increases, whereas it decreases as $\theta$ increases. This behavior agrees with theory since wall pressure fluctuations in the low-frequency range are caused by larger turbulent structures that scale with $\delta$, while fluctuations in the high-frequency range relate to small turbulent structures that scale with the viscous length ($\nu/u_\tau$) which has a direct relationship with $\tau_w$. In addition, the inflow velocity ($U$) increases the WPS level throughout the frequency range, as expected. Furthermore, the inflow velocity slightly shifts the energy content towards higher frequencies (see the shift in the maximum in Fig.~\ref{subfig:sens_u}), which is also in agreement with the theory~\cite{Devenport}.

To obtain a single equation that models the WPS in the entire frequency range, the WPS for each band are linearly weighted with the frequency, as:
\begin{equation} \label{eq:freq_pond}
\begin{split}
\Phi_\mathrm{pp|LMF} =&\Phi_\mathrm{pp|LF} \frac{\log_{10}(f_3) - \log_{10}(f)}{\log_{10}(f_3) - \log_{10}(f_1)} + \Phi_\mathrm{pp|MF}\frac{\log_{10}(f) - \log_{10}(f_1)}{\log_{10}(f_3) - \log_{10}(f_1)}; \\
\Phi_\mathrm{pp|MHF} =& \Phi_\mathrm{pp|MF} \frac{\log_{10}(f_4) - \log_{10}(f)}{\log_{10}(f_4) - \log_{10}(f_2)} + \Phi_\mathrm{pp|HF}  \frac{\log_{10}(f) - \log_{10}(f_2)}{\log_{10}(f_4) - \log_{10}(f_2)}; \\
\Phi_\mathrm{pp} = &\Phi_\mathrm{pp|LMF} \frac{\log_{10}(f_4) - \log_{10}(f)}{\log_{10}(f_4) - \log_{10}(f_1)} + \Phi_\mathrm{pp|MHF} \frac{\log_{10}(f) - \log_{10}(f_1)}{\log_{10}(f_4) - \log_{10}(f_1)}.
\end{split}
\end{equation}
Initially, the low- and mid-frequency bands are combined between $f_1$ and $f_3$; and the mid- and high-frequency bands are combined between $f_2$ and $f_4$. In this process, we obtain two curves, defined as $\Phi_\mathrm{pp|LMF}$ and $\Phi_\mathrm{pp|MHF}$, respectively. Later, these two expressions are combined using the same approach as between $f_1$ and $f_4$.

\subsection{Error assessment between the experiments and the symbolic regression model}
The error is measured as the square difference between the predicted ($\Phi_\mathrm{pp|mod}$) and measured ($\Phi_\mathrm{pp|meas}$) WPS (in dB), weighted by the frequency to account for the logarithmic scale, as:
\begin{equation}\label{eq:error}
   \mathrm{Error} = \frac{1}{\log \left({\frac{f_\mathrm{max}}{f_o}} \right)} \int_{f_o}^{f_\mathrm{max}} \frac{1}{f} \left( \Phi_\mathrm{pp|meas} -  \Phi_\mathrm{pp|mod}\right)^2~\mathrm{d}f;
\end{equation}
where $\log$ is the natural logarithm, and $f_o$ and $f_\mathrm{max}$ are the minimum and maximum frequencies where the overall spectrum is calculated.

\begin{figure}[H]
	\centering
	\begin{subfigure}[c]{0.45\textwidth}
		\centering
		\includegraphics[width=\textwidth]{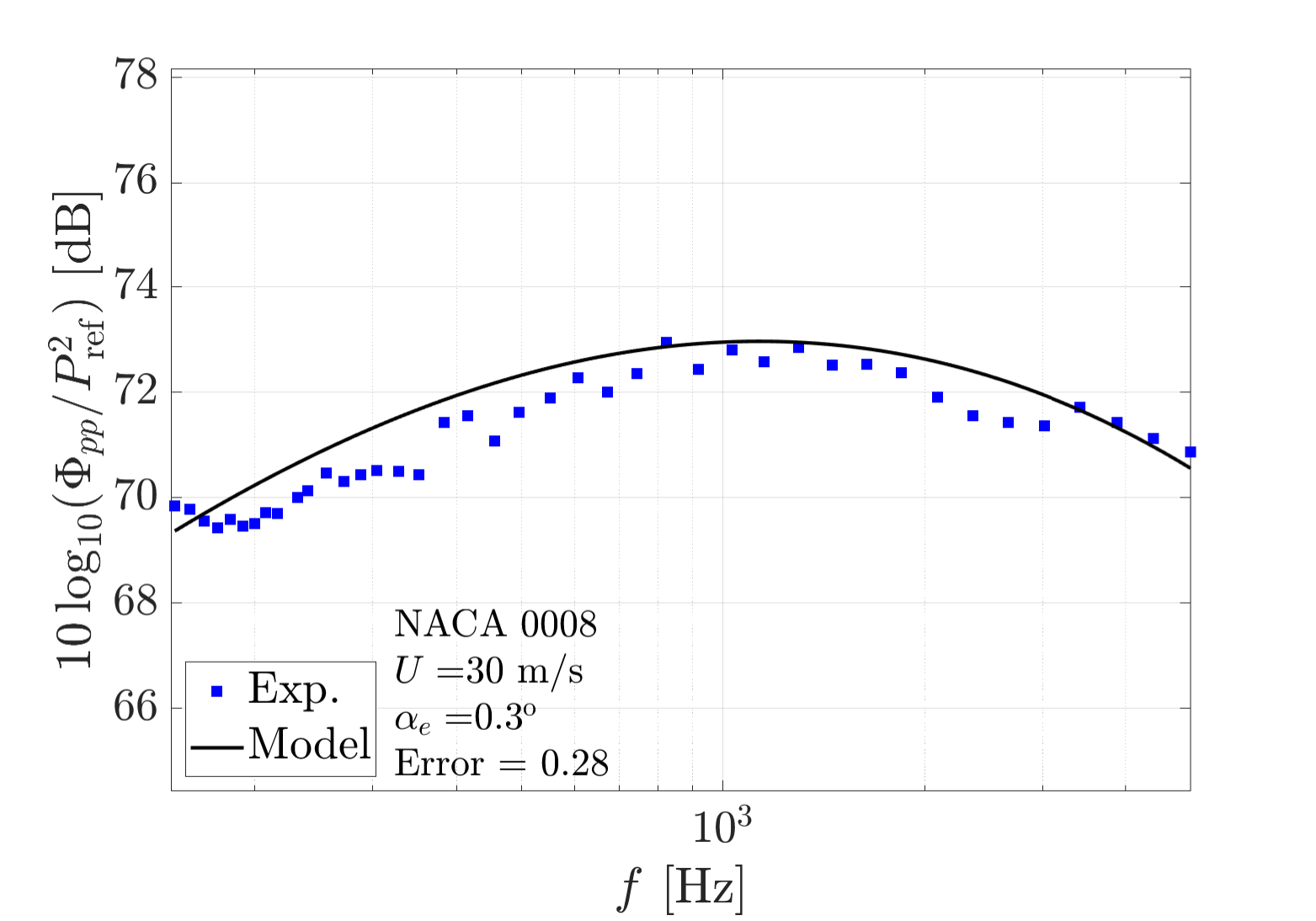}
		\label{subfig:val_a}
	\end{subfigure}
	\hfill
	\begin{subfigure}[c]{0.45\textwidth}
		\centering
		\includegraphics[width=\textwidth]{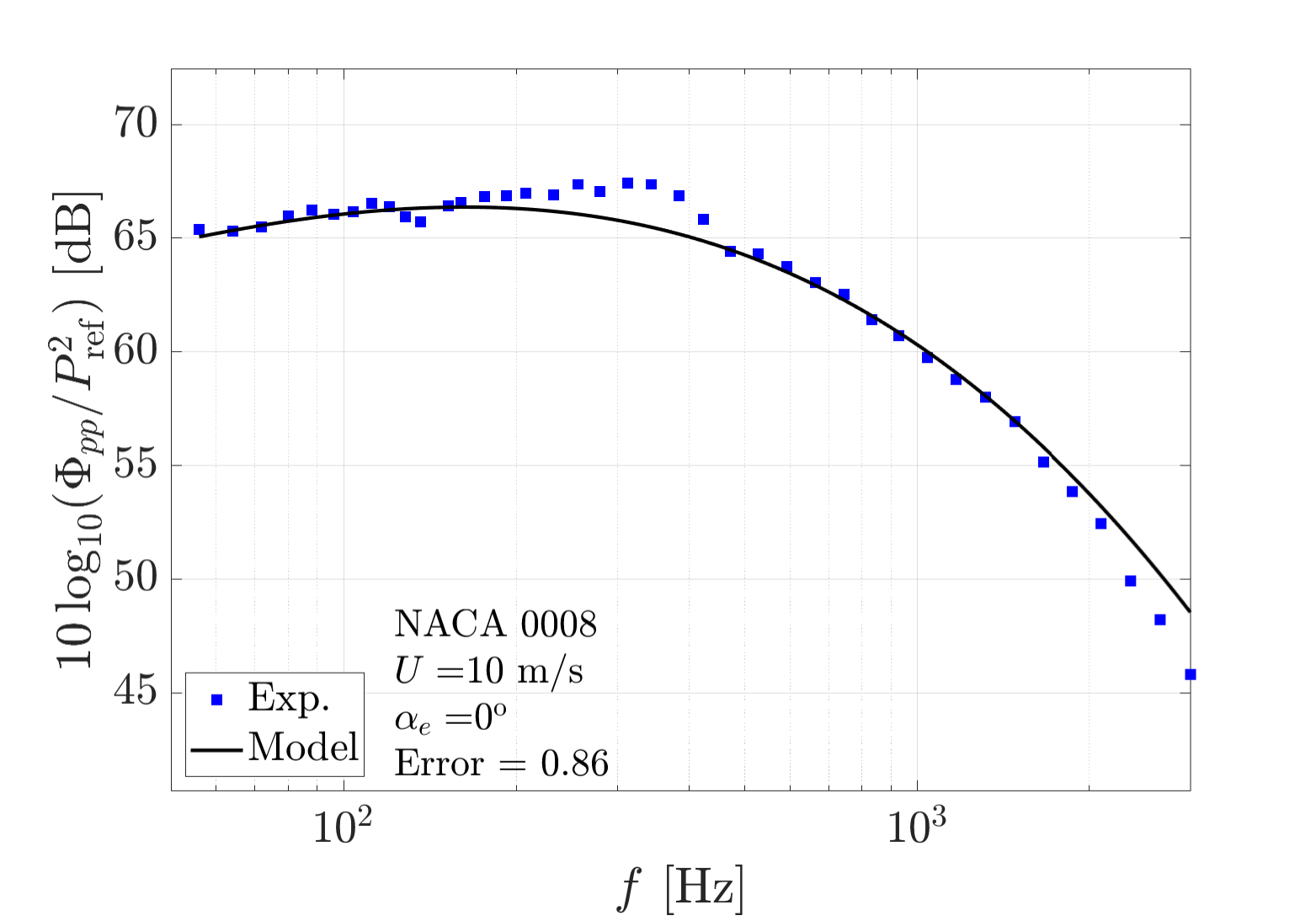}
		\label{subfig:val_b}
	\end{subfigure}	
	\begin{subfigure}[c]{0.45\textwidth}
		\centering
		\includegraphics[width=\textwidth]{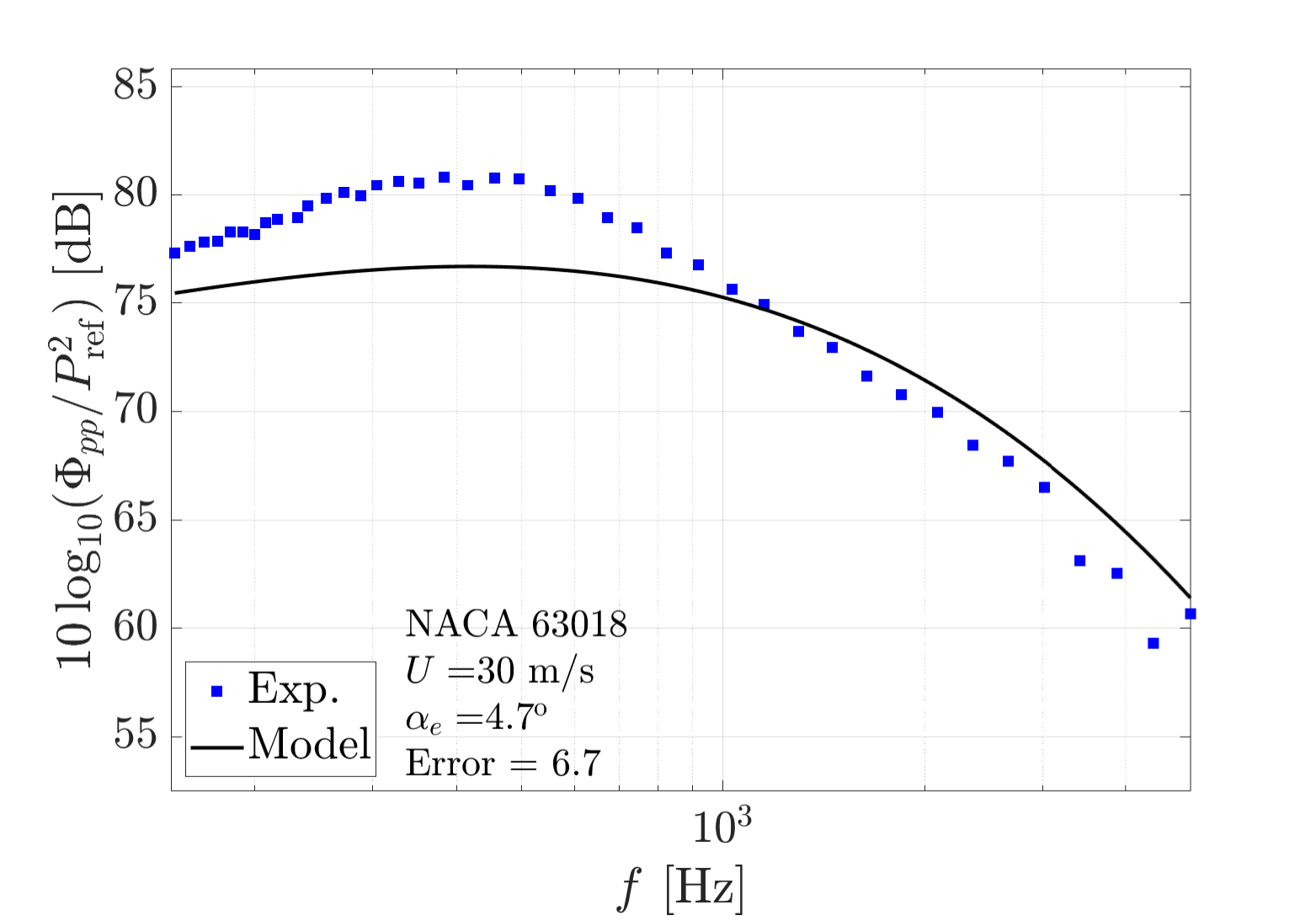}
		\label{subfig:val_c}
	\end{subfigure}
	\hfill
	\begin{subfigure}[c]{0.45\textwidth}
		\centering
		\includegraphics[width=\textwidth]{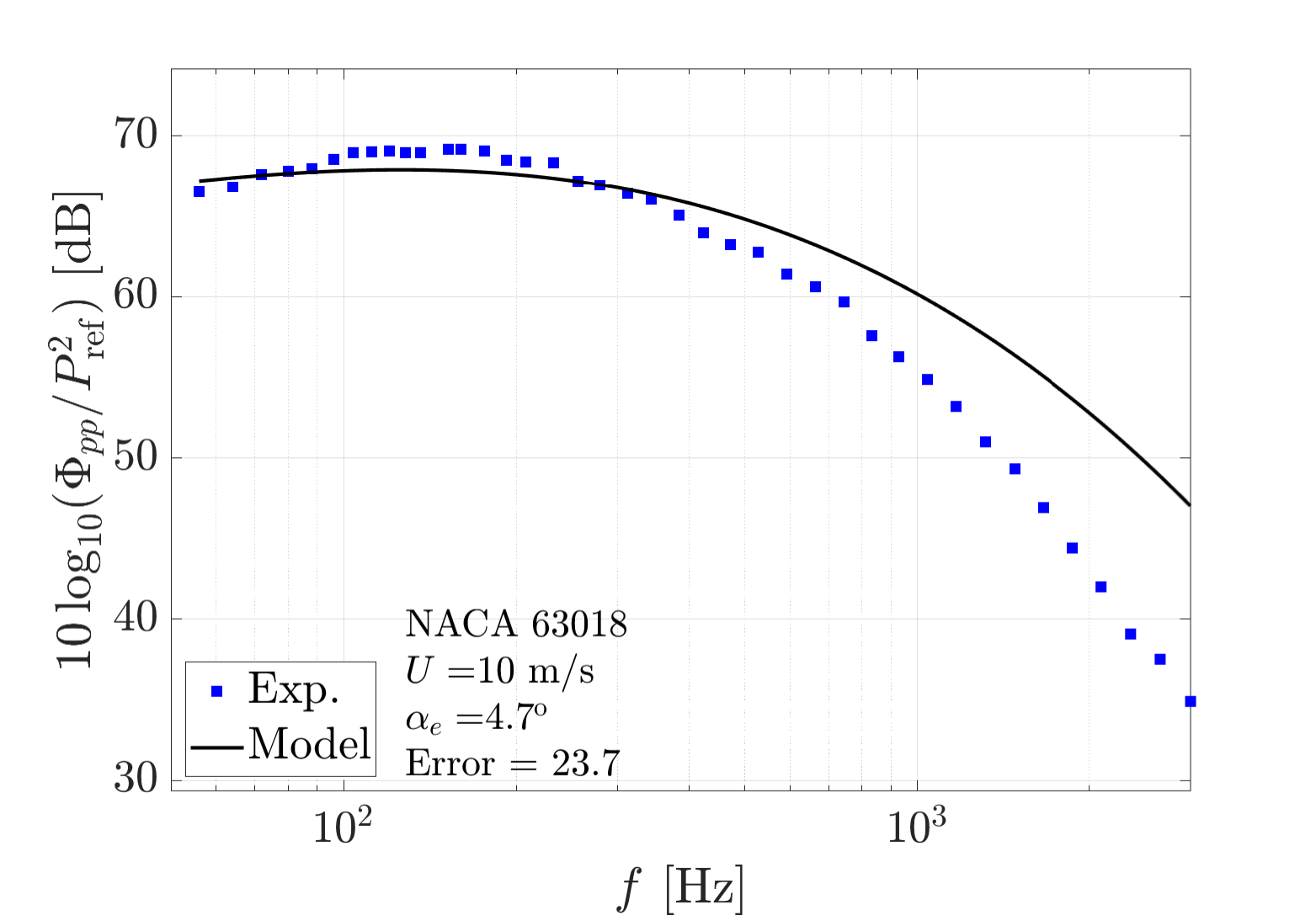}
		\label{subfig:val_d}
	\end{subfigure}
	\caption{Predicted and measured WPS of several cases from the training dataset.
    }
	\label{fig:error_dataset}
\end{figure}

Table~\ref{tab:error_training} shows the error statistics of the model for the training dataset calculated with Eq.~\ref{eq:error}. Note that the error is segregated into favorable (FPG) and adverse (APG) pressure gradients; and between low-frequency (LF) and high-frequency (HF) bands, since the size of the dataset was different, ref. to Section~\ref{sec:dataset}. The error is similar for the four groups of cases, showing slightly better results for the APG (lower mean and maximum values). For completeness, in Fig.~\ref{fig:error_dataset} the measured and predicted WPS for four cases within the dataset are shown, with the associated errors that range from the minimum to maximum (see Table~\ref{tab:error_training}). The precision of the model slightly decreases with the angle of attack; however, it agrees with the experimental results in the entire frequency range with a margin of 4~dB. 
\begin{table}[h!]
    \centering
    \begin{tabular}{|c|c|c|c|c|}
    \hline
       \diagbox{Error}{Cases} & APG \& LF & FPG \& LF  & APG \& HF & FPG \& HF \\
        \hline
      Error\textsubscript{min}  & 0.24   & 0.17& 0.49 & 0.45\\
      Error\textsubscript{max}  & 23.88  &11.56 & 8.32 &11.56\\
      Error\textsubscript{mean}  & 3.24  &2.09 & 2.66 &2.44\\
       \hline
    \end{tabular}
    \caption{\label{tab:error_training}Error of the model for the training dataset.}
\end{table}

\section{Validation of the model outside the training dataset}\label{sec:val}
The model is validated by comparing the predictions with published experimental data and classic semi-empirical models. Later, acoustic predictions for a full-scale wind turbine are also presented and compared to field measurements. 

We challenge the model with predictions outside the range of the training dataset and explore how it generalizes for Reynolds numbers, angle of attack, airfoil geometry, and frequency ranges outside the range of the training set. Now, the Reynolds number range is \{0.26 - 1.5\}$\cdot$10\textsuperscript{6} compared to  \{0.26 - 0.90 \}$\cdot$10\textsuperscript{6} of the training data set. The maximum angle of attack is 10.3\textsuperscript{o} compared to 7\textsuperscript{o} in the dataset. More importantly, three different airfoil geometries are used for validation: NACA~0012, NACA~0018, and DU-96-W-180. Furthermore, experiments were conducted in a different experimental setup. 

\begin{figure}[H]
	\centering
	\begin{subfigure}[c]{0.45\textwidth}
		\centering
		\includegraphics[width=\textwidth]{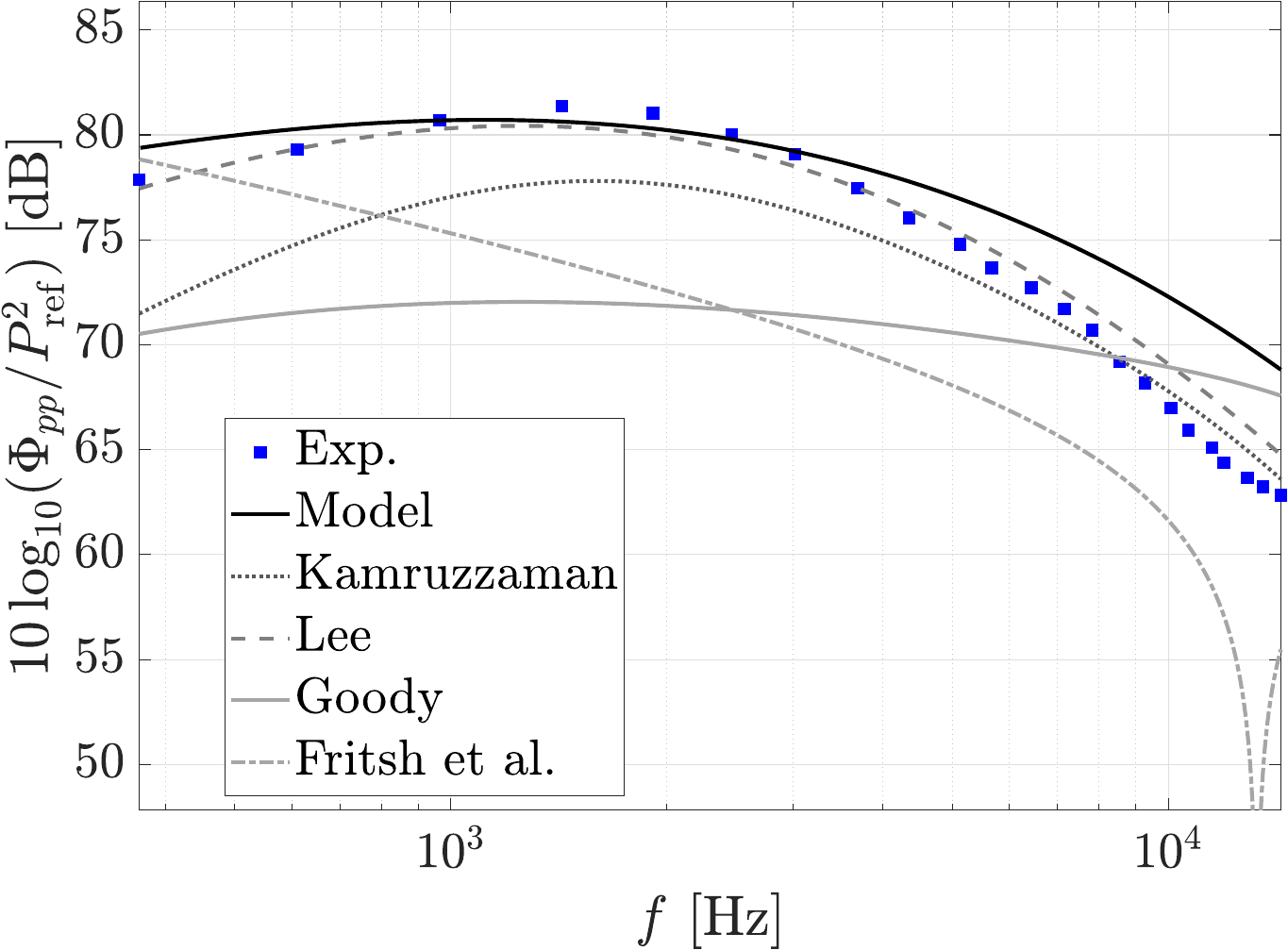}
		\caption{Case A}
		\label{subfig:case_a}
	\end{subfigure}
	\hfill
	\begin{subfigure}[c]{0.45\textwidth}
		\centering
		\includegraphics[width=\textwidth]{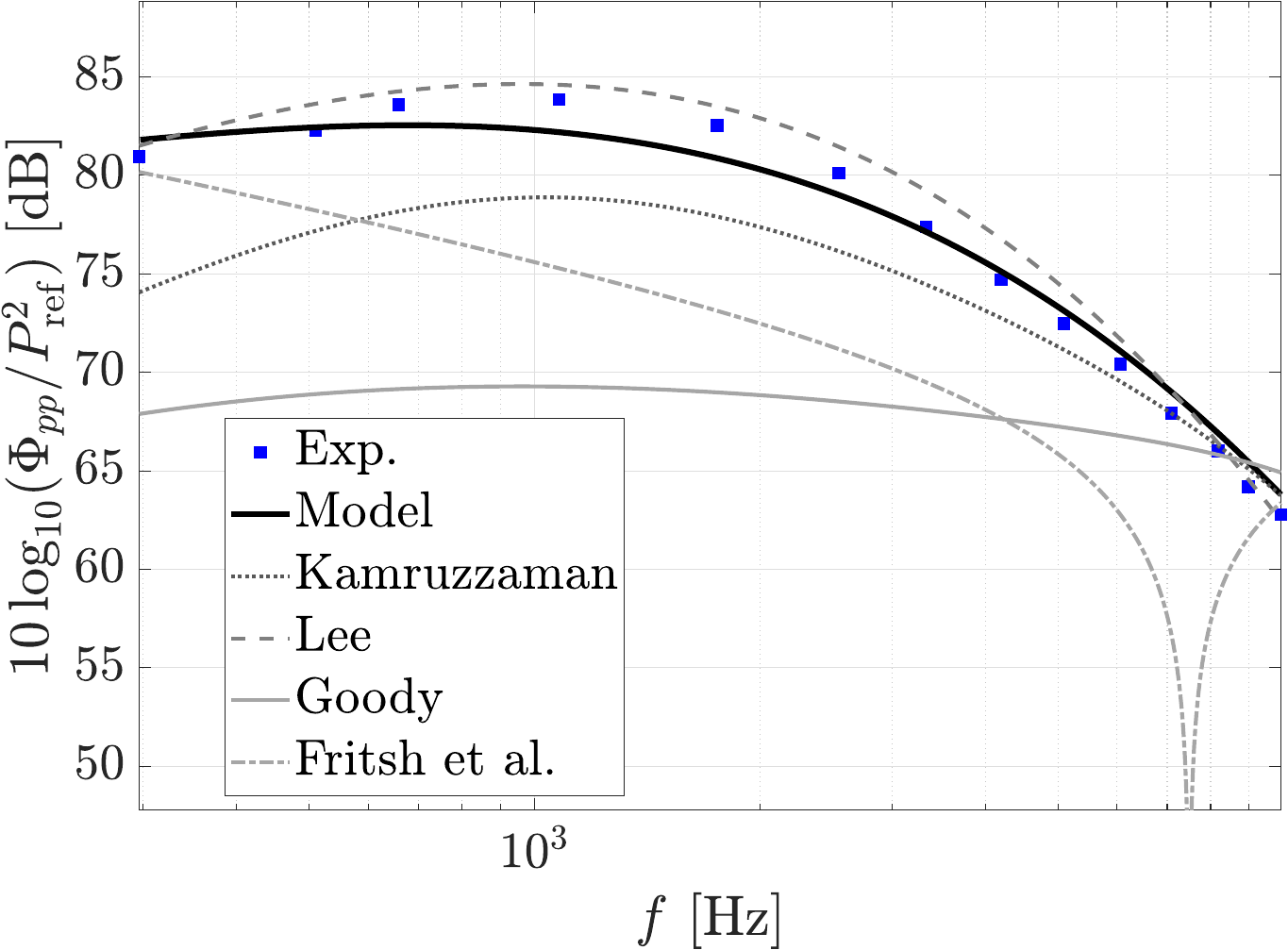}
		\caption{Case B}
		\label{subfig:case_b}
	\end{subfigure}	
	\begin{subfigure}[c]{0.45\textwidth}
		\centering
		\includegraphics[width=\textwidth]{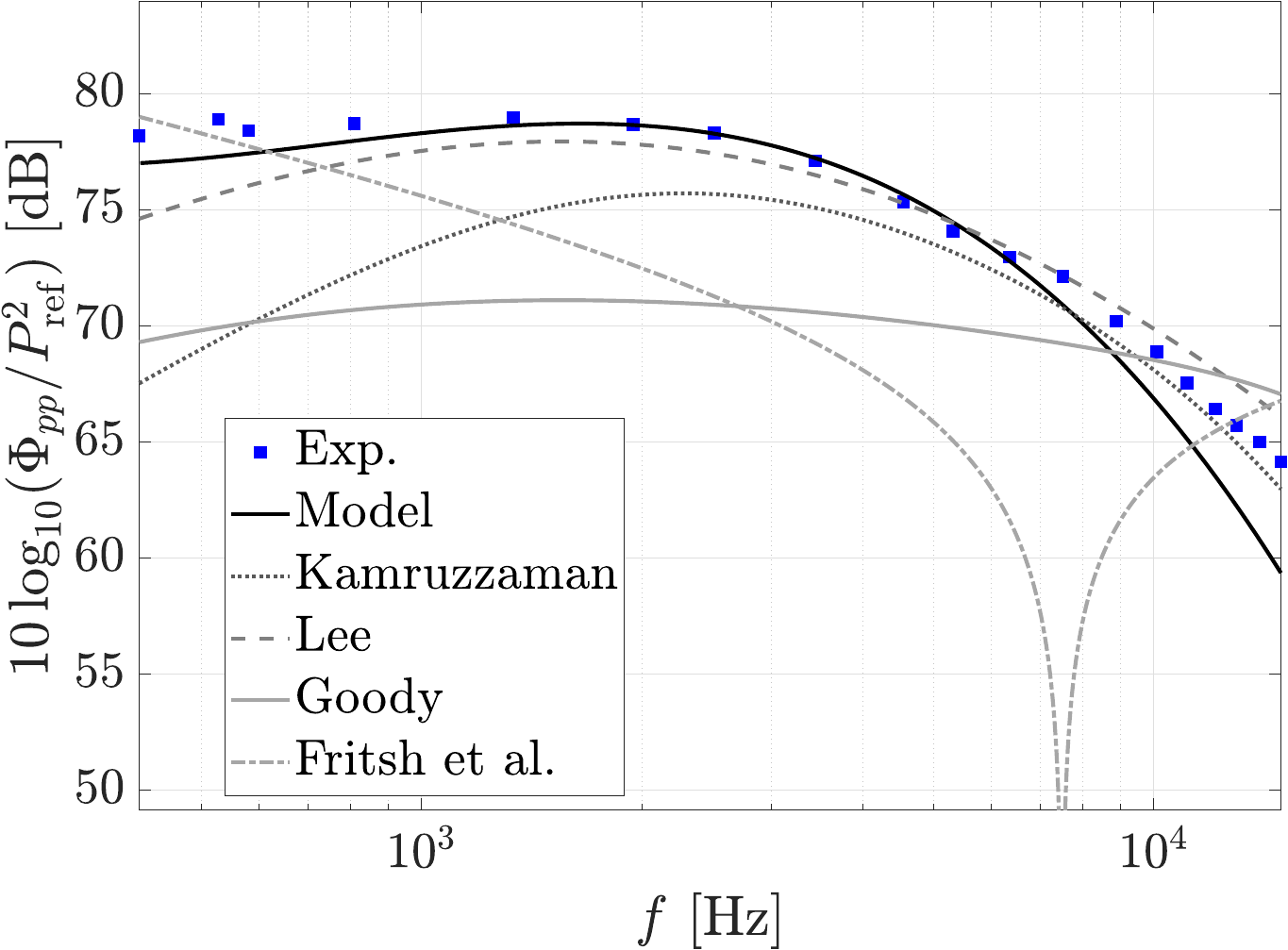}
		\caption{Case C}
		\label{subfig:case_c}
	\end{subfigure}
	\hfill
	\begin{subfigure}[c]{0.45\textwidth}
		\centering
		\includegraphics[width=\textwidth]{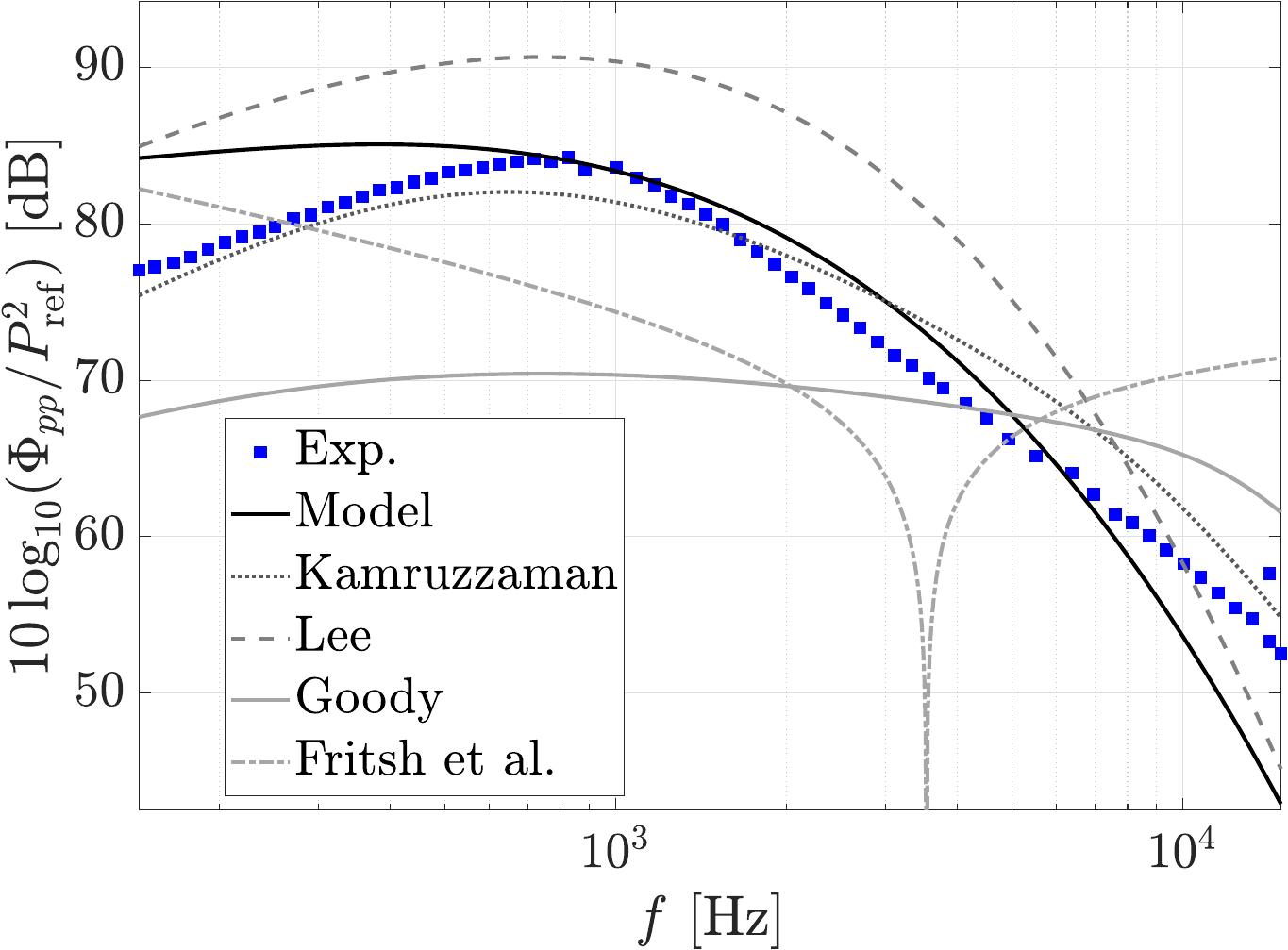}
		\caption{Case D}
		\label{subfig:case_d}
	\end{subfigure}
    \begin{subfigure}[c]{0.45\textwidth}
		\centering
		\includegraphics[width=\textwidth]{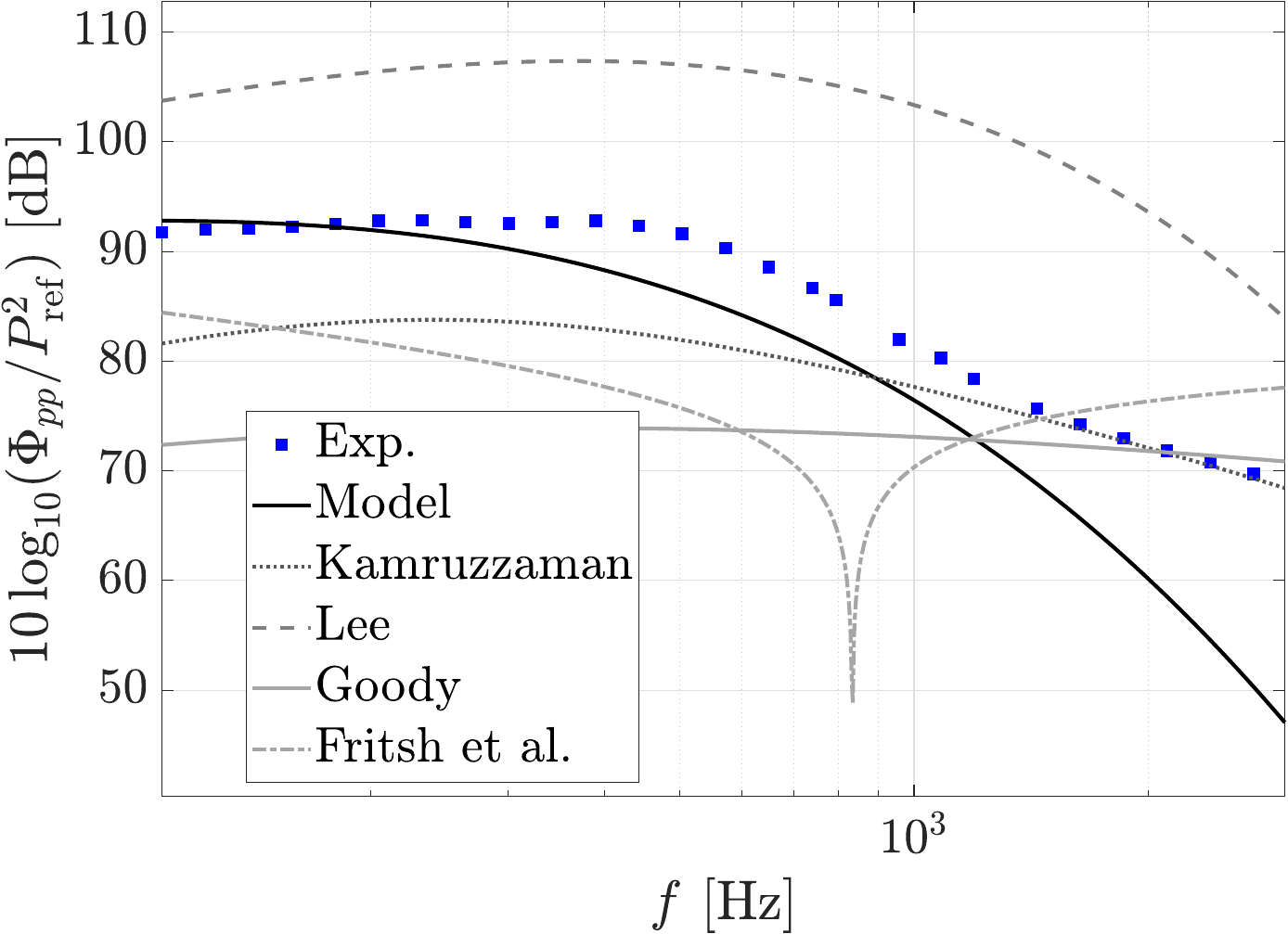}
		\caption{Case E}
		\label{subfig:case_e}
	\end{subfigure}
    \hfill
	\begin{subfigure}[c]{0.45\textwidth}
		\centering
		\includegraphics[width=\textwidth]{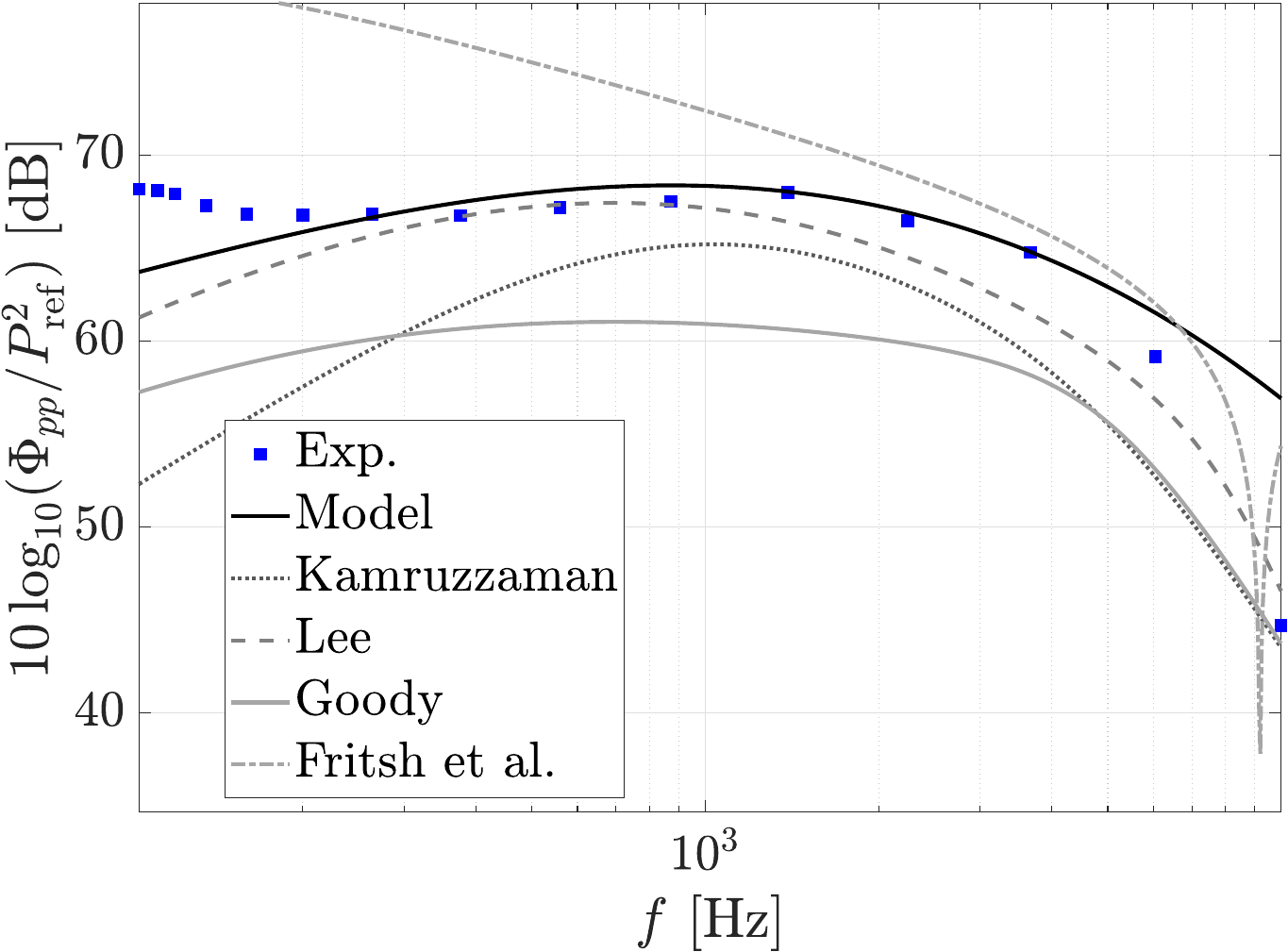}
		\caption{Case F}
		\label{subfig:case_f}
	\end{subfigure}
\begin{subfigure}[c]{0.45\textwidth}
		\centering
		\includegraphics[width=\textwidth]{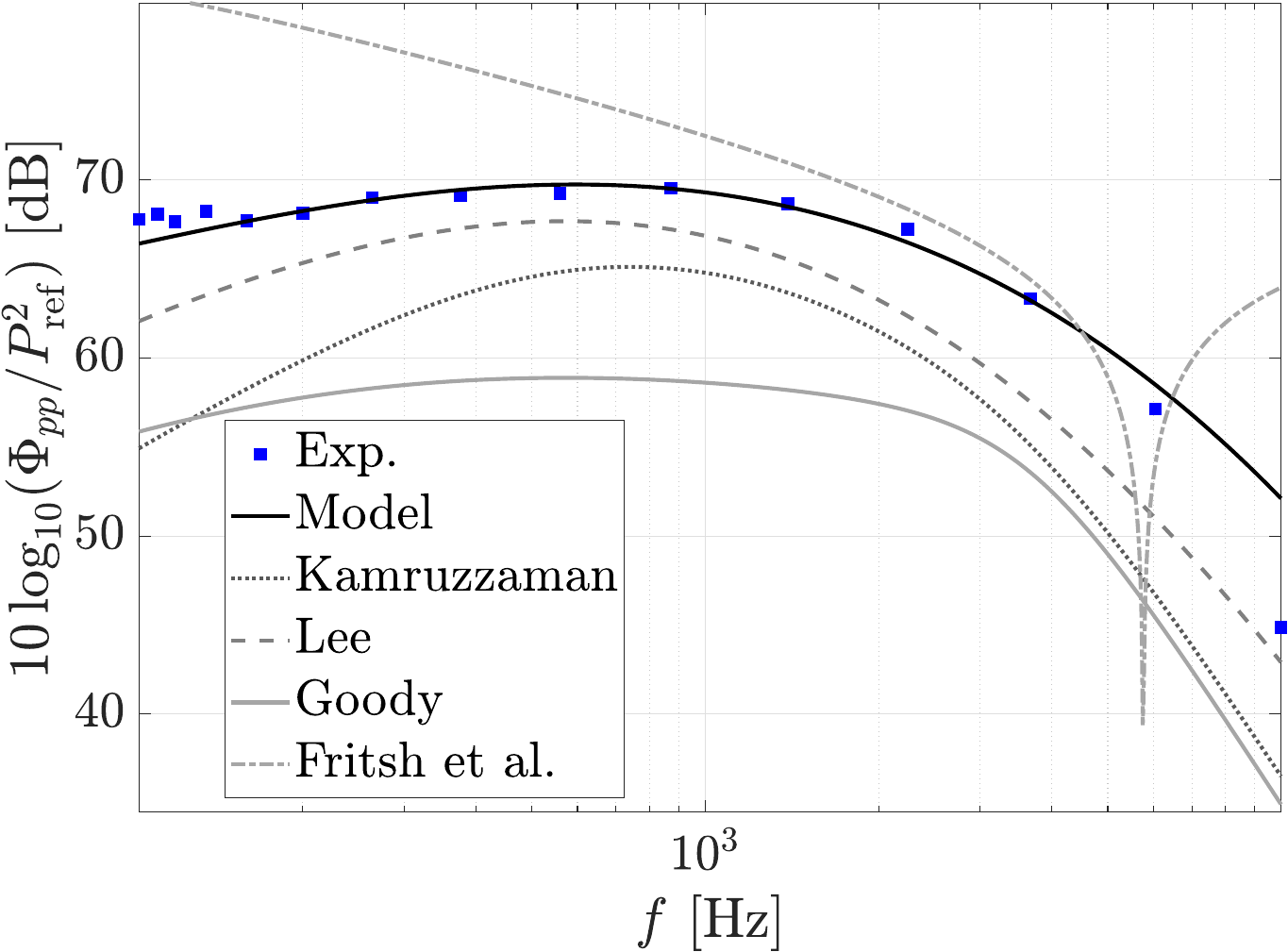}
		\caption{Case G}
		\label{subfig:case_g}
\end{subfigure}
    \hfill
	\begin{subfigure}[c]{0.45\textwidth}
		\centering
		\includegraphics[width=\textwidth]{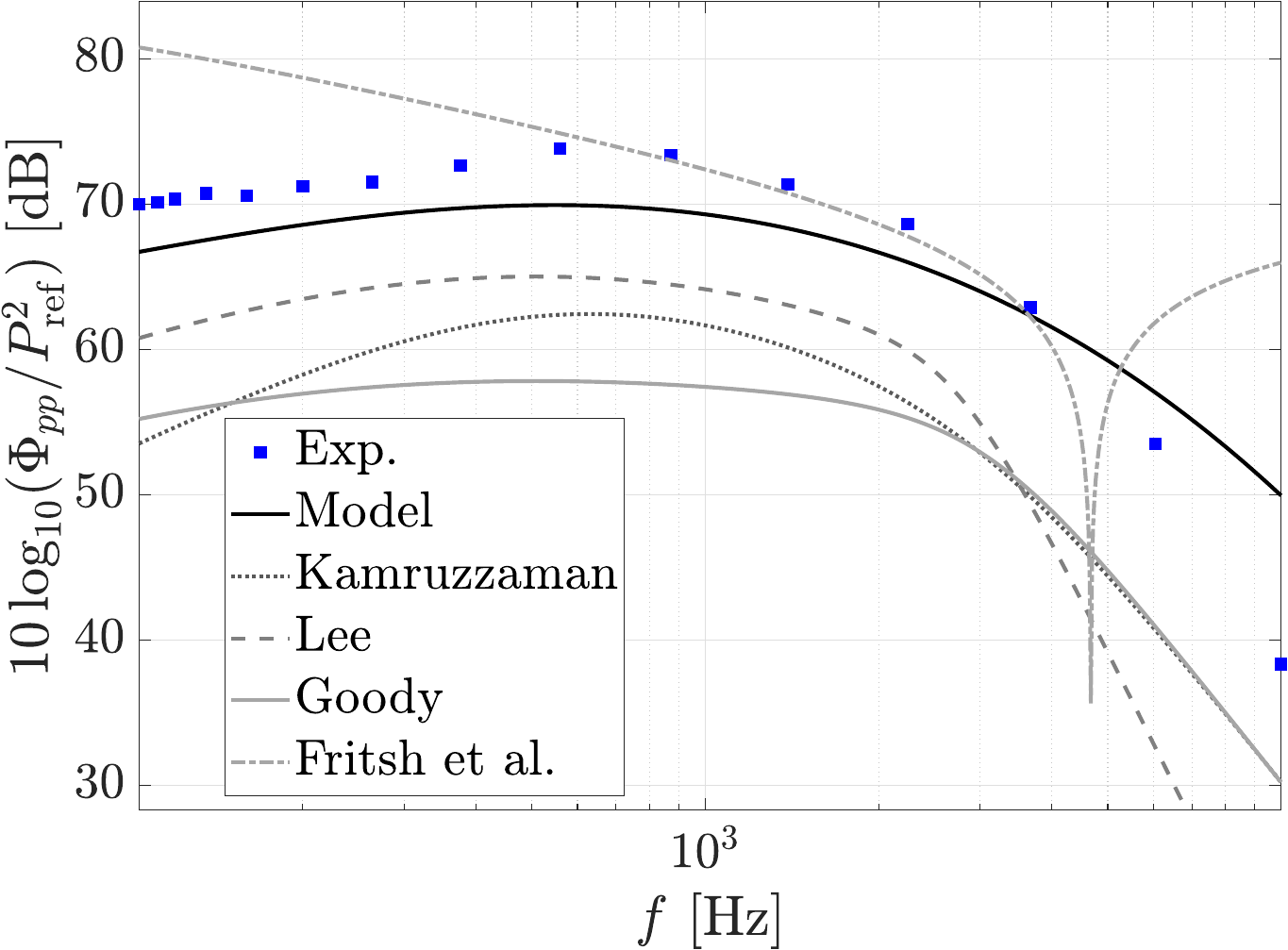}
		\caption{Case H}
		\label{subfig:case_h}
	\end{subfigure}
	\caption{Comparison of the predicted and measured WPS for several conditions outside the training dataset.}
	\label{fig:validation}
\end{figure}

Figure~\ref{fig:validation} shows the measured WPS (Exp. in the legend) compared with our model (model in the legend), together with the Kamruzzaman model~\cite{kamruzzaman2015semi}, Lee model~\cite{lee2017}, Goody model~\cite{goody2004}, and the symbolic regression-based model proposed by \citet{fritsch2023modeling}. The specific conditions for each validation case are summarized in Table~\ref{tab:validation}, together with the error (calculated following Eq.~\ref{eq:error}), and the reference of the experimental data. 
\begin{table}[h!]
\footnotesize
    \centering
    \begin{tabular}{|c|c|c|c|c|c|c|}
    \hline
        \multirow{2}{*}{Name} &  \multirow{2}{*}{Conditions}  & \multicolumn{4}{c|}{Error} & \multirow{2}{*}{Ref.}  \\
        \cline{3-6}
       & & \small{Kamruzzaman} &\small{Lee} & \small{Goody} & \small{Model} & \\
        \hline
       \multirow{2}{*}{Case A}  & NACA 0012; $U_\infty$=56;& \multirow{2}{*}{11.54} &  \multirow{2}{*}{\textbf{1.05}} & \multirow{2}{*}{49.33} & \multirow{2}{*}{6.60} & \multirow{2}{*}{\cite{kuccukosman2018}}\\
        &  $c$=0.4; $\alpha$=0\textsuperscript{o}&  &  &  &  & \\
        \hline
        \multirow{2}{*}{Case B}  &  NACA 0012; $U_\infty$=56;& \multirow{2}{*}{19.61} & \multirow{2}{*}{1.43} & \multirow{2}{*}{137.00} & \multirow{2}{*}{\textbf{1.26}} & \multirow{2}{*}{\cite{kuccukosman2018}}\\
        & $c$=0.4; $\alpha$=4 \textsuperscript{o}; SS&  &  &  &  & \\
        \hline
        \multirow{2}{*}{Case C}  & NACA 0012; $U_\infty$=56; & \multirow{2}{*}{22.83} & \multirow{2}{*}{2.30} & \multirow{2}{*}{43.64} & \multirow{2}{*}{\textbf{1.77}} & \multirow{2}{*}{\cite{kuccukosman2018}}\\
        & $c$=0.4; $\alpha$=4\textsuperscript{o}; PS&  &  &  &  & \\
        \hline
        \multirow{2}{*}{Case D}  &DU-96-W-180; $U_e$=56; & \multirow{2}{*}{\textbf{5.87}} & \multirow{2}{*}{59.83} & \multirow{2}{*}{87.81} & \multirow{2}{*}{14.16} & \multirow{2}{*}{\cite{DU96}}\\
        & $c$=0.3; $\alpha$=4.6\textsuperscript{o}; SS &  &  &  &  & \\
        \hline
        \multirow{2}{*}{Case E}  &DU-96-W-180; $U_e$=66; & \multirow{2}{*}{37.11} & \multirow{2}{*}{294.92} & \multirow{2}{*}{221.08} & \multirow{2}{*}{\textbf{31.38}} & \multirow{2}{*}{\cite{DU96}}\\
        & $c$=0.3; $\alpha$=10.3\textsuperscript{o}; SS &  &  &  &  & \\
        \hline
        \multirow{2}{*}{Case F}  & NACA 0012; $U_\infty$=20;& \multirow{2}{*}{41.43 } & \multirow{2}{*}{\textbf{5.39}} & \multirow{2}{*}{44.94} & \multirow{2}{*}{7.35} & \multirow{2}{*}{\cite{sanders2022}}\\
        &  $c$=0.2; $\alpha$=0\textsuperscript{o}&  &  &  &  & \\
        \hline
        \multirow{2}{*}{Case G}  & NACA 0018; $U_\infty$=20; & \multirow{2}{*}{60.18} & \multirow{2}{*}{16.79} & \multirow{2}{*}{113.86} & \multirow{2}{*}{\textbf{1.81}} & \multirow{2}{*}{\cite{sanders2022}}\\
        &$c$=0.2; $\alpha$=0\textsuperscript{o}&  &  &  &  & \\
        \hline
        \multirow{2}{*}{Case H}  & NACA 63018; $U_\infty$=20;& \multirow{2}{*}{149.06} & \multirow{2}{*}{149.45} & \multirow{2}{*}{193.70} & \multirow{2}{*}{\textbf{14.62}} & \multirow{2}{*}{\cite{sanders2022}}\\
        &  $c$=0.2; $\alpha$=0\textsuperscript{o}&  &  &  &  & \\
        \hline
        \multicolumn{2}{|c|}{Average} & 43.44 & 66.39 & 111.41& \textbf{9.86} &\\
        \cline{1-6}
        \multicolumn{2}{|c|}{Maximum} & 149.06 & 294.92 & 221.08& \textbf{31.38} &\\
        \cline{1-6}
        \multicolumn{2}{|c|}{$\sigma$} & 46.14 & 105.52 & 68.53& \textbf{10.18} &\\
        \hline
    \end{tabular}
    \caption{\label{tab:validation}Error of several models compared to experimental measurements outside the training dataset.}
\end{table}

In general, the proposed model shows good agreement for all cases. The best agreement is obtained for cases B, C, and G, with errors lower than the average error of the training dataset. The worst case is for case~E ($\alpha$=10.3\textdegree), where there is a significant discrepancy between the model and measurements for $f>$1~kHz. However, our model predicts well the low-frequency hump expected for high angles of attack~\cite{DU96, botero2023_eif_rmp}. The main reason for the poor agreement in the high-frequency range is that $\tau_w$, which significantly affects the spectrum rolloff in the high frequency (see Fig.\ref{subfig:sens_tau_w}), is obtained with XFOIL simulations instead of experimental measurements as the other parameters ($\delta, \theta, \delta^*$) that are reported in \citet{DU96}. The reference~\cite{DU96} also shows that the precision of XFOIL simulations at high angles of attack is reduced. Even though case~E is the worst case for our model, it presents the lowest error among all models, which shows that the model is robust enough and can be extrapolated even for complex scenarios. 

Compared with the other semi-empirical models, our model also performs well. There are some specific cases where other approaches present a better agreement with the measurements than our model, such as, Lee~model for cases A and F, and Kamruzzaman~model for case D. However, our model presents the lowest average error among all the models. Furthermore, our model is the most robust since the error variability is the lowest among the models and the maximum error is also the lowest. In particular, the model proposed by \citet{fritsch2023modeling} does not show good behavior for the particular cases presented here, mainly because of the divergence at high frequencies. This frequency divergence was avoided in our model because of the constraints imposed in the training process of the symbolic regression algorithm.

\newpage

\begin{figure}[h]
\centering
\includegraphics[width=0.5\textwidth]{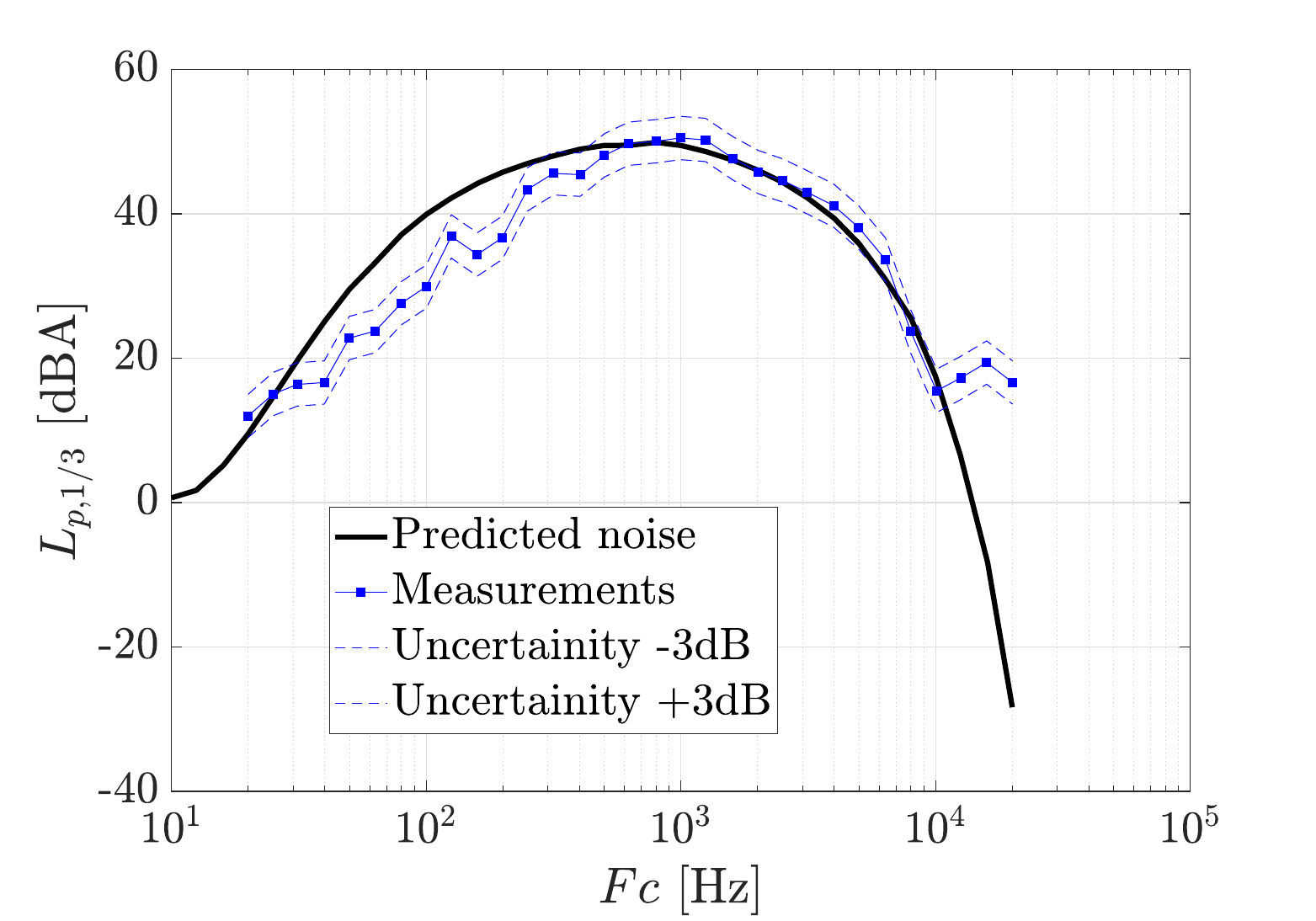}
\caption{
Wind turbine noise prediction and measurements using the proposed WPS model.}\label{fig:wt_noise}
\end{figure}

Finally, the trailing-edge noise produced by a SWT-93-2.3 wind turbine is predicted using the model proposed in this research to calculate the WPS on the suction and pressure sides. Noise prediction also includes turbulence-interaction noise. The results are compared with the measurements reported in \citet{zephyr_database}. The noise prediction approach is reported in \citet{botero2024_wt_noise}, together with the aerodynamic performance of the turbine. The operating conditions of the wind turbine are: 17.5~rpm rotational speed, 9.5~m/s inflow velocity, and 5\textdegree~pitch angle. The observer is located on the ground, 100~m downstream the wind turbine. The blade conditions (airfoil geometry, chord, twist, angle of attack, and apparent velocity), where the WPS is calculated, are shown in~\ref{app:wt_blade}. In this final validation, the WPS model is extrapolated to different airfoil geometries, inflow velocities, and angle of attack. There is a good agreement between the predicted noise and the measurements (see Fig.~\ref{fig:wt_noise}), mainly in the frequency range where trailing-edge noise is more relevant, i.e., $f>$0.7~kHz~\cite{botero2024_wt_noise}. The results show once again the robustness of the model and its good performance for an extensive range of applications. 

\section{Conclusions}\label{sec:con}
This paper presents a new semiempirical model for predicting the WPS of turbulent boundary layers of mid- to high-Reynolds numbers under both adverse and favorable pressure gradients. The model is derived from experimental measurements of wall-pressure fluctuations on the suction and pressure sides of two airfoils across a wide range of inflow velocities and angles of attack. Unlike conventional single-equation models, the proposed model divides the WPS into three frequency bands, accounting for the physics of the turbulent scales that contribute to wall pressure fluctuations in different frequency ranges. Symbolic regression is used to derive mathematical expressions for each frequency range, using boundary layer parameters that can be obtained from simple tools such as XFOIL (or CFD simulations), thus preserving the simplicity and efficiency of single-equation models. The resulting model consists of five equations: three for the WPS levels in each frequency range and two for the boundary frequencies.

Within the training dataset, the model predicts the wall-pressure spectrum with a maximum integrated error of 23~dB, which is a difference in the spectral level of 4~dB.

The model is validated against experimental data from the ldifference of maximum 4~d of the wall-pressure spectrum level between the measured and predicted data. iterature that encompass various airfoils, Reynolds numbers, and angles of attack. The results demonstrate the good performance of the model under a variety of conditions. Slight reduction in accuracy is observed for strong adverse pressure gradients (pre-stall angles of attack) and is attributed to XFOIL predictions at high angles of attack. Overall, the model exhibits greater robustness compared to widely used semi-empirical and empirical alternatives, since it has the lowest maximum and average error and the lowest standard deviation for the validation cases. The standard deviation of our model is 10.2, compared to 46.1, 68.5, and 105.5 observed for other semiempirical methods. 

Finally, the model is challenged to predict full-scale wind turbine trailing-edge noise, achieving good agreement (within less than 1~dB) with field measurements.

\section*{Acknowledgments}
The authors would like to acknowledge Ing. W. Lette, ir. E. Leusink, S. Wanrooij for the technical support on the experimental setup and conduction of the experiments.

\section*{Funding Sources}
This research has received funding from the European Union (ERC, Off-coustics, project number 101086075) and also from the European Commission through the H2020-MSCA-ITN-209 project zEPHYR (grant agreement No 860101). 
Views and opinions expressed are, however, those of the authors only and do not necessarily reflect those of the European Union or the European Research Council. Neither the European Union nor the granting authority can be held responsible for them.
DH and EF acknowledge the funding received by the Comunidad de Madrid according to Orden 5067/2023, of December 27th, issued by the Consejero de Educación, Ciencia y Universidades, which announces grants for the hiring of predoctoral research personnel in training for the year 2023.

\appendix
\section{Remote microphone probes calibration}\label{sec:rmp_cal}
\begin{figure}[htb!]
    \centering
        \includegraphics[width=0.49\textwidth]{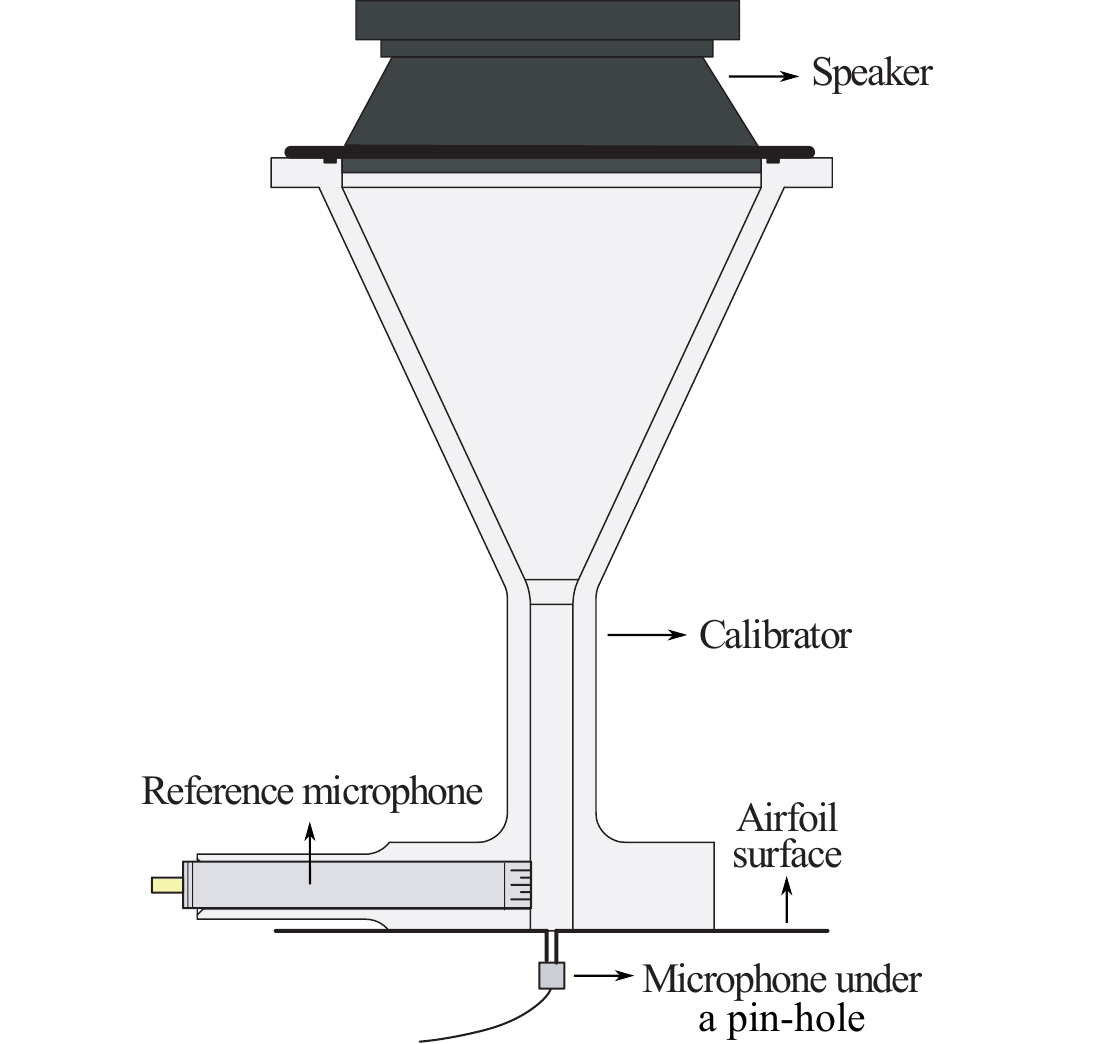}
        \caption{\label{Fig: calibrator} In-house calibrator.} 
\end{figure}

The calibration of the remote microphone probes consisted of obtaining a transfer function between an ideal microphone that is flushed-mounted with the airfoil surface and the remote microphone probe. For that, an in-house calibrator, shown in Fig.~\ref{Fig: calibrator} is used. The calibrator design is based on that presented by \citet{Roger2017}. It is equipped with a reference microphone, i.e., a GRAS 40HP, and an FR8 loudspeaker. The noise source was white noise. The calibration was done in situ when the airfoils were installed in the wind tunnel and consisted of two steps to minimize the influence of the calibrator in the calibration procedure~\cite{Roger2017,berntsen2014remote}. The first step consisted in measuring the reference microphone (ref) in the calibrator and a microphone flush-mounted (FM), i.e., a GRAS 40HP, simultaneously. This results in the transfer function:
 \begin{equation}\label{eq: transfer_function}
     \mathrm{TF}_{\mathrm{ref,FM}} = \frac{\Phi_{\mathrm{ref,ref}}}{\Phi_{\mathrm{ref,FM}}}.
 \end{equation}
Then, a transfer function between the reference microphone in the calibrator and the remote microphone probe (RMP), i.e., a Knowles FG 23329-P07, was obtained:
 \begin{equation}\label{eq: transfer_function2}
    \mathrm{TF}_{\mathrm{ref, RMP}} = \frac{\Phi_{\mathrm{ref,ref}}}{\Phi_{\mathrm{ref,RMP}}}.
 \end{equation}
The transfer function between the microphone flush-mounted and the remote microphone probe is the ratio of the two transfer functions:
 \begin{equation}\label{eq: transfer_function3}
   \lvert \mathrm{TF}_{\mathrm{FM, rmp}} \rvert ^2 = \frac{\lvert \mathrm{TF}_{\mathrm{ref, RMP}} \lvert ^2}{\lvert \mathrm{TF}_{\mathrm{ref, FM}} \lvert ^2}.
 \end{equation}
In Eqs.~\ref{eq: transfer_function} to~\ref{eq: transfer_function3}, $\Phi_{\mathrm{x,x}}$ and $\Phi_\mathrm{{x,y}}$ are the auto- and cross-spectrum of microphone signals represented by the subindex in the equations. The measurements of the RMP were corrected using the final transfer function, i.e., Eq.~\ref{eq: transfer_function3}, to obtain the equivalent spectrum of the WPS at the airfoil surface($\Phi_{pp}$):
 \begin{equation}\label{eq: phi_eq}
     \Phi = \frac{\Phi_{\mathrm{rmp, rmp}}}{\lvert TF_{\mathrm{FM, rmp}} \rvert ^2}
 \end{equation}

The coherence between the reference microphone in the calibrator and the remote microphone probe was analyzed to ensure that both microphones were in the same acoustic field. Due to contamination in the high-frequency range of electronic noise during the calibration, the spectrum is analyzed only up to a frequency of 5~kHz.

\section{XFOIL simulations validation}\label{sec:app_Xfoil_val}
XFOIL simulations were conducted to obtain the boundary layer parameters used in the symbolic regression approach to propose the model. The validation is performed by comparing the results with measurements of the boundary layer and pressure coefficient distribution. This appendix addresses the methodology of the measurements of the boundary layer and steady surface pressure and the comparison between experimental and XFOIL results. 

\subsection{Boundary layer measurements}
Boundary layer measurements were performed at 97\% of the airfoil chord on the suction side, using a Dantec Dynamics model 55P15 single-wire probe of 5~\textmu m diameter and 1.25~mm wire length. The hot wire data were acquired with the Dantec StreamLine Pro CTA system coupled with a National Instruments 9215 A/D converter. Temperature correction was conducted for the measurements, adopting 21\textdegree C as the reference temperature. Measurements were acquired during 20~s with a sampling frequency of 65536 (2\textsuperscript{16})~Hz. An anti-aliasing cutoff filter was used at a frequency of 30~kHz. Furthermore, a high-pass filter of 10~Hz was used during the processing of the data to eliminate the effects related to the flow buffeting instability that is naturally present in open wind tunnels in the test section~\cite{Heesen2004Windtunnel}. The probe was mounted in a Dantec Dynamics 55H22 probe support installed on a symmetric airfoil, which was fixed in a 3D traverse system, allowing probe translation with a resolution of 6.5~\textmu m. 
 
The hot-wire calibration was performed in situ in the close test section with a Prandtl tube as a reference. The calibration consisted of 32 velocity points distributed logarithmically ranging from 2.5 to 50~m/s. The velocity measurements had a maximum system uncertainty of 5\% with a confidence interval of 95\%. This uncertainty was computed following the guidelines provided by Dantec Dynamics, which considers calibration equipment, calibration linearization, A/D board resolution, probe positioning, and temperature variations.

 On average, velocity was measured at 35 locations across the boundary layer and five points in the free stream. The distribution of the measurements across the boundary layer was logarithmic. The distance of the probe to the wall was determined by the contact of a feeler gauge to the hot-wire prongs. The gauge accuracy is 0.05 mm. The distance from the wall of the first measurement varied from 0.5~mm to 1~mm among the configurations. 
 
 The experimental boundary layer displacement thickness ($\delta^*$) and momentum thickness ($\theta$) are determined by performing a trapezoidal numerical integration of the measured boundary layer velocity profile. The boundary layer thickness ($\delta$) is determined according to Eq.~\ref{eq:delta_xfoil}, following the same approach as that used in the XFOIL simulations. The friction velocity is calculated by fitting the experimental mean velocity profile to the Prandtl-von Kármán log-law coupled with Coles' wake law: 
\begin{equation}\label{eq: mean_velocity_profile}
\frac{U}{u_\tau} = \frac{1}{\kappa}\log \left( y^+ \right) + B + \frac{2\Pi_w}{\kappa}\sin^2\left( \frac{\pi y}{2\delta} \right).
\end{equation}
The fitting also determines the wake factor $\Pi_w$. The parameter $\kappa$ denotes the von Kármán constant, equal to 0.38, B is a level constant equal to 5, and $y^+$ is the non-dimensional distance from the wall, defined as $y^+ = y u_\tau/\nu$.
 
\subsection{Steady surface pressure measurements}
The anechoic termination of the remote microphone probes located along the chord was connected to a NetScanner model 9216 pressure scanner to obtain the static pressure on the airfoil surface. Therefore, the unsteady and steady surface pressure were measured simultaneously. Twenty sensors are located on the suction and pressure sides distributed along the chord. Measurements were taken during 30~s with a sampling frequency of 300~Hz. The pressure coefficient, $C_p$, is calculated according to: 
\begin{equation}\label{Eq: Cp}
C_\mathrm{p} = \frac{p-P_\infty}{\frac{1}{2} \rho U^2},
\end{equation}
where $\rho$ is the air density, and U is the inflow velocity, which is measured by a Prandtl tube located at the end of the close test section. 

 \subsection{XFOIL validation with experimental data}
\begin{figure}[hbt!]
 \begin{centering}
    \begin{tabular}{cc}
{\footnotesize{}\includegraphics[width=.5\textwidth]{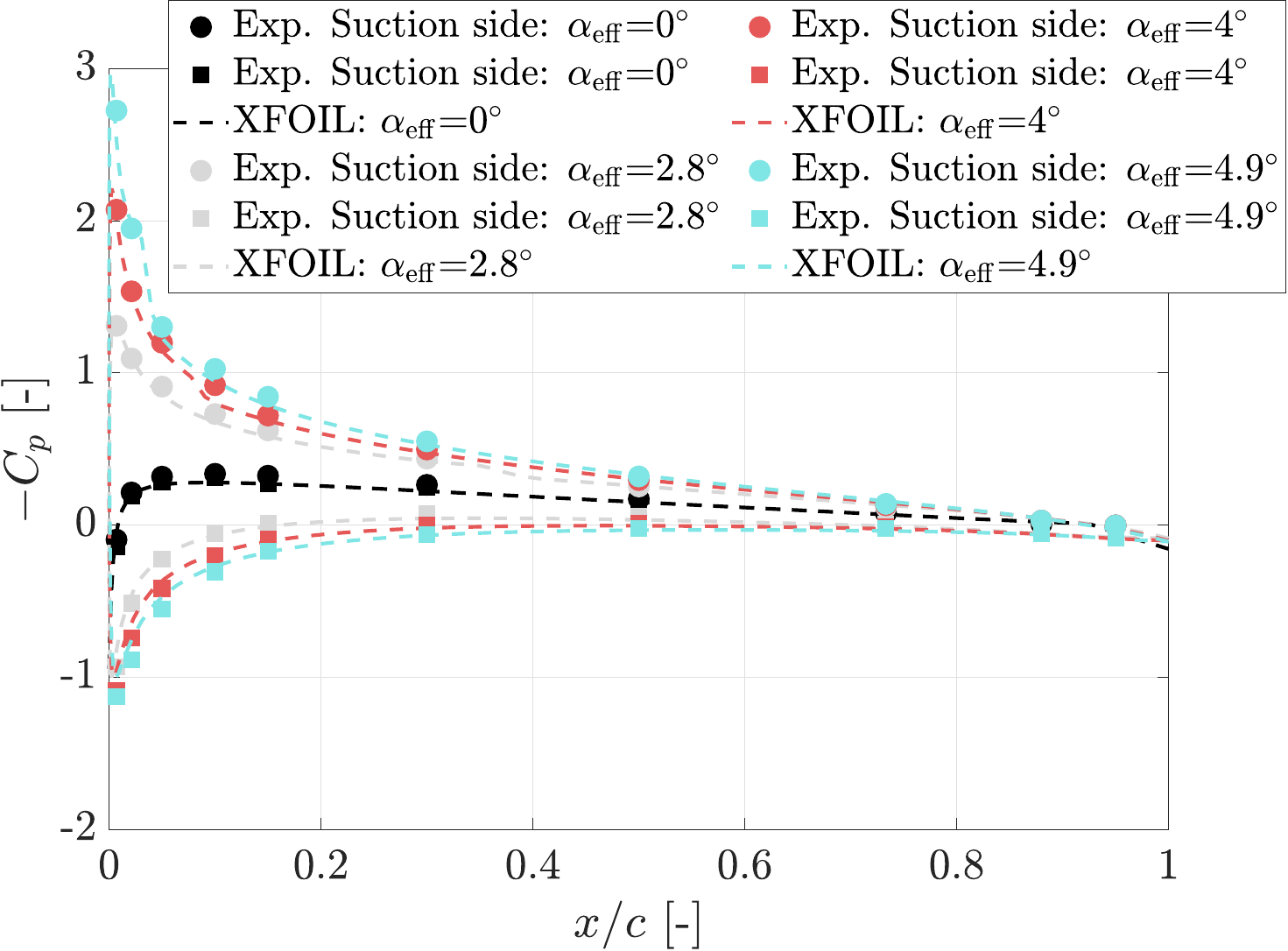}} &
{\footnotesize{}\includegraphics[width=.5\textwidth]{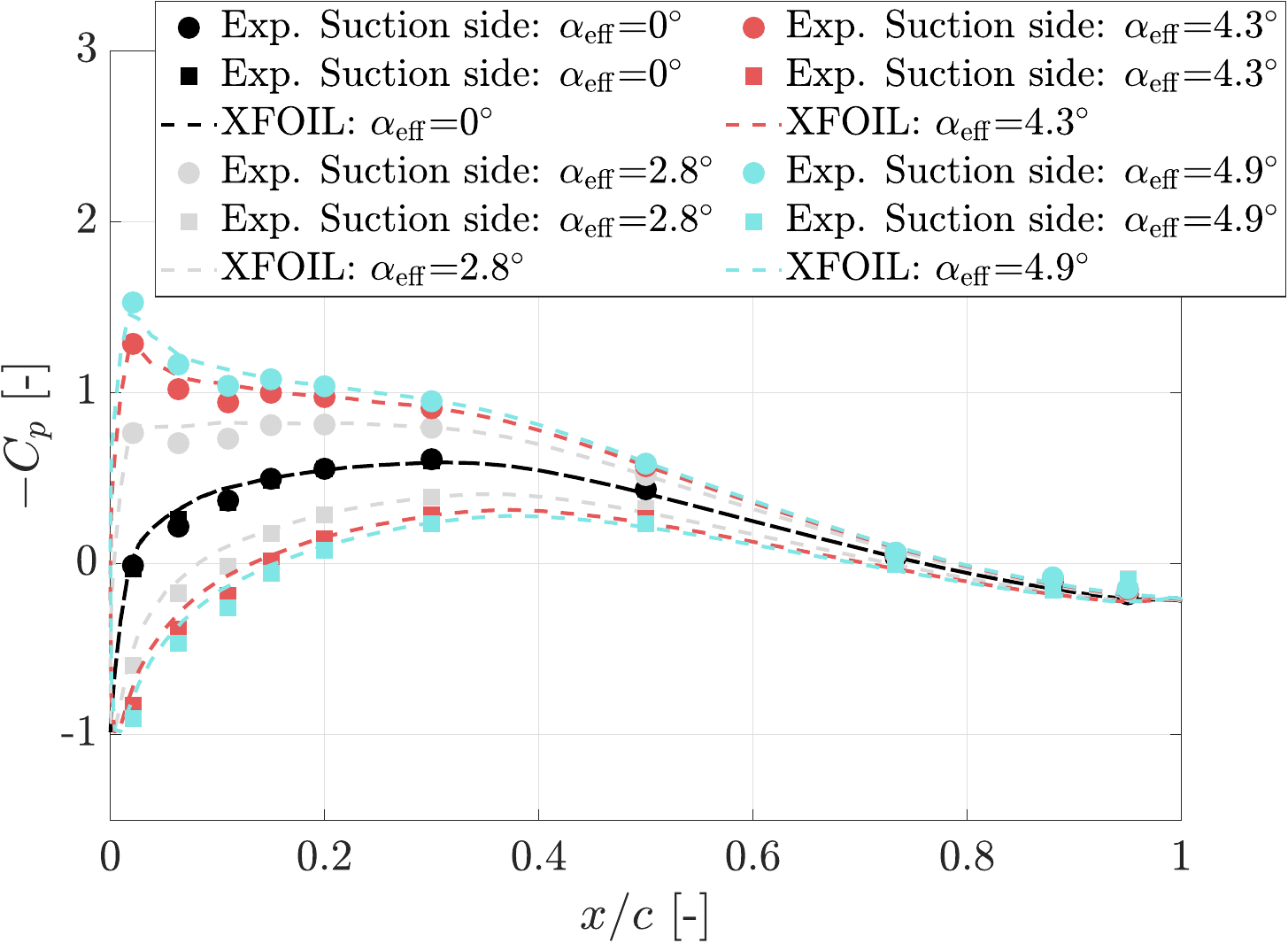}}
\tabularnewline
{\footnotesize{} (a) NACA~0008} & {\footnotesize{} (b) NACA~63018}
    \end{tabular}
\caption{\label{Fig:Cp} Pressure coefficient distribution compared with XFOIL results for several angles of attack. $U~=~30$ m/s. $Re~=~600\times 10^3$.}
\par\end{centering}
\end{figure}

Figure~\ref{Fig:Cp} shows the comparison of the pressure coefficient ($Cp$) distribution obtained experimentally and with XFOIL for $U$=30~m/s at different effective angles of attack for both airfoils. A good agreement is obtained in all the cases for both, suction and pressure sides. XFOIL captures well the suction peak and the pressure gradient along the chord. A small deviation is obtained between $x/c$=0.065 and 0.1 because those pressure ports were located right before and after the tripping device, which disturbs the mean flow. 

Tables~\ref{tab:delta_s_NACA0008} to~\ref{tab:delta_NACA63018} show the comparison between experimental measurements and XFOIL simulations of $\delta^*$, and $\theta$ and calculated $u_\tau$ and $\delta$. There is a good agreement between the XFOIL results and the experimental data for airfoils and all conditions. XFOIL follows the same trend as the experimental results with the velocity and angle of attack, such as $\delta$ increasing with the angle of attack and decreasing with the inflow velocity, while $u_\tau$ decreasing with the angle of attack. The highest error appears for higher angles of attack and lower velocities. However, the errors are acceptable considering the many factors that can influence the boundary layer measured, such as the determination of the distance from the wall and the tripping device and transition process, which are more critical for lower Reynolds numbers. In general, using XFOIL to estimate boundary layer parameters is reasonable to propose the WPS model, given that it is a common methodology to predict WPS and trailing-edge noise~\cite{Stalnov2016, Lee_TE_Xfoil, bertagnolio_TE_Xfoil}. 
\begin{table}
\centering
    \begin{tabular}{|c|c|c|c|}
    \hline
   \multirow{2}{*}{ $U$ [m/s]} &\multicolumn{3}{c|}{$\delta^*$ [mm]}
 \\ \cline{2-4}
     & Exp. & XFOIL &  $\frac{\vert\mathrm{Exp.} - \mathrm{XFOIL\vert}}{\mathrm{Exp.}}$ \\ \hline
     10 & \{1.5, 2.3, 4.0\} & \{1.9, 2.5, 3.3\} & \{26.6\%, 8.7\%, 17.5\% \}\\
     30 &\{1.1, 1.8, 2.8\} &\{1.4, 1.9, 2.4\} & \{ 27.7\%, 5.5\%, 14.3\% \}  \\
     45& \{-, 1.7, 2.5\} & \{-, 1.7, 2.2\} &  \{0.5\%, 12.0\% \}  \\ \hline 
    \end{tabular}
    \caption{Comparison between XFOIL and experimental measurements of $\delta^*$ on the suction side for different angles of attack for the NACA~0008 airfoil. The values between curly brackets correspond to the boundary layer parameters for $\alpha_\mathrm{eff}=$~\{0\textdegree, 3\textdegree, 5\textdegree\}.}
    \label{tab:delta_s_NACA0008}
    \end{table}
   \begin{table}
\centering
        \begin{tabular}{|c|c|c|c|}
    \hline
   \multirow{2}{*}{ $U$ [m/s]} & \multicolumn{3}{c|}{$\theta^*$ [mm]} 
 \\ \cline{2-4}
     & Exp. & XFOIL & $\frac{\vert\mathrm{Exp.} - \mathrm{XFOIL\vert}}{\mathrm{Exp.}}$\\ \hline
     10  &\{1.1, 1.4, 2.1\} &\{1.2, 1.5, 1.9\} & \{9.1\%, 7.1\%, 9.5\%\}\\
     30 &\{0.8, 1.2, 1.6\}& \{0.9, 1.2 1.5\} & \{ 12.5\%, 0.5\%, 6.2\%\}  \\
     45& \{-, 1.1 1.6\} & \{-, 1.1, 1.4\} & \{0.3\%, 12.5\% \} \\ \hline 
    \end{tabular}
    \caption{Comparison between XFOIL and experimental measurements of $\theta^*$ on the suction side for different angles of attack for the NACA~0008 airfoil. The values between curly brackets correspond to the boundary layer parameters for $\alpha_\mathrm{eff}=$~\{0\textdegree, 3\textdegree, 5\textdegree\}.}
    \label{tab:theta_s_NACA0008}
    \end{table}
   \begin{table}
\centering
    \begin{tabular}{|c|c|c|c|}
    \hline
   \multirow{2}{*}{ $U$ [m/s]} &\multicolumn{3}{c|}{$u_\tau$ [m/s]}  \\ \cline{2-4}
    & Exp. & XFOIL &  $\frac{\vert\mathrm{Exp.} - \mathrm{XFOIL\vert}}{\mathrm{Exp.}}$  \\ \hline
     10 & \{0.37, 0.29, 0.17\} & \{0.39, 0.36, 0.33\} & \{-, 5.4\%, 24.1\%, 9.41\%\}\\
     30 & \{1.08, 0.88, 0.62\} & \{1.1, 1.02, 0.68\} & \{-, 1.9\%, 15.9\%, 9.7\%\} \\
     45& \{-, 1.31, 1.34\}  &\{-, 1.5, 1.39\} & \{-, 14.5\%, 3.7\%\} \\ \hline 
    \end{tabular}
    \caption{Comparison between XFOIL and experimental measurements of $u_\tau$ on the suction side for different angles of attack for the NACA~0008 airfoil. The values between curly brackets correspond to the boundary layer parameters for $\alpha_\mathrm{eff}=$~\{0\textdegree, 3\textdegree, 5\textdegree\}.}
    \label{tab:u_tau_NACA0008}
    \vspace{5 mm}
    \begin{tabular}{|c|c|c|c|}
    \hline
   \multirow{2}{*}{ $U$ [m/s]} & \multicolumn{3}{c|}{$\delta$ [mm]}  \\ \cline{2-4}
   & Exp. & XFOIL &  $\frac{\vert\mathrm{Exp.} - \mathrm{XFOIL\vert}}{\mathrm{Exp.}}$ \\ \hline
     10 &\{8.9, 11.3, 14.7\}&\{9.2, 11.4, 14.3\} & \{-, 3.3\%, 0.9\%, 2.7\%\}\\
     30 & \{7.0, 9.3, 12.1\} &\{7.4, 9.1, 11.4\}& \{-, 5.7\%, 2.1\%, 5.8\%\}\\
     45&\{-, 9.6, 12.7\}& \{-, 8.4, 10.5\} & \{-, 12.5\%, 17.3\%\}\\ \hline 
    \end{tabular}
    \caption{Comparison between XFOIL and experimental measurements of $\delta$ on the suction side for different angles of attack for the NACA~0008 airfoil. The values between curly brackets correspond to the boundary layer parameters for $\alpha_\mathrm{eff}=$~\{0\textdegree, 3\textdegree, 5\textdegree\}.}
    \label{tab:delta_NACA0008}
\end{table}

\begin{table}
    \centering
    \begin{tabular}{|c|c|c|c|}
    \hline
       \multirow{2}{*}{ $U$ [m/s]} &\multicolumn{3}{c|}{$\delta^*$ [mm]} 
 \\ \cline{2-4}
     & Exp. & XFOIL &  $\frac{\vert\mathrm{Exp.} - \mathrm{XFOIL\vert}}{\mathrm{Exp.}}$\\ \hline
     10 & \{3.3, 3.7\} & \{3.4, 4.7\} &\{3.0\%, 27.0\% \}\\
     30 & \{2.0, 2.1\} &\{2.5, 3.4\}&\{25.0\%, 61.9\%  \}\\
    \hline 
    \end{tabular}
    \caption{Comparison between XFOIL and experimental measurements of $\delta^*$ on the suction side for different angles of attack for the NACA~63018 airfoil. The values between curly brackets correspond to the boundary layer parameters for $\alpha_\mathrm{eff}=$~\{0\textdegree, 3\textdegree\}.}
    \label{tab:delta_s_NACA63018}
   \end{table}
   \begin{table}
\centering
        \begin{tabular}{|c|c|c|c|}
    \hline
       \multirow{2}{*}{ $U$ [m/s]} & \multicolumn{3}{c|}{$\theta^*$ [mm]} 
 \\ \cline{2-4}
     & Exp. & XFOIL &  $\frac{\vert\mathrm{Exp.} - \mathrm{XFOIL\vert}}{\mathrm{Exp.}}$\\ \hline
     10  &\{1.6, 1.7\}& \{1.8, 2.2\}&\{12.5\%, 29.4\%  \}\\
     30  &\{1.3, 1.4\}&\{1.5, 1.8\}&\{ 15.4\%, 28.5\%\}\\
    \hline 
    \end{tabular}
    \caption{Comparison between XFOIL and experimental measurements of $\theta^*$ on the suction side for different angles of attack for the NACA~63018 airfoil. The values between curly brackets correspond to the boundary layer parameters for $\alpha_\mathrm{eff}=$~\{0\textdegree, 3\textdegree\}.}
    \label{tab:theta_s_NACA63018}
     \end{table}
   \begin{table}
\centering
    \begin{tabular}{|c|c|c|c|}
    \hline
   \multirow{2}{*}{ $U$ [m/s]} &\multicolumn{3}{c|}{$u_\tau$ [m/s]} \\ \cline{2-4}
     & Exp. & XFOIL &  $\frac{\vert\mathrm{Exp.} - \mathrm{XFOIL\vert}}{\mathrm{Exp.}}$  \\ \hline
     10 &\{0.22, 0.14\} & \{0.28, 0.22\} &\{27.3\%, 57.1\%\} \\
     30 &\{0.83, 0.64\} &\{0.81, 0.68\} &\{2.4\%, 6.3\% \}\\
    \hline 
    \end{tabular}
    \caption{Comparison between XFOIL and experimental measurements of $u_\tau$ on the suction side for different angles of attack for the NACA~63018 airfoil. The values between curly brackets correspond to the boundary layer parameters for $\alpha_\mathrm{eff}=$~\{0\textdegree, 3\textdegree\}.}
    \label{tab:u_tau_NACA63018}
     \end{table}
   \begin{table}
\centering
    \begin{tabular}{|c|c|c|c|}
    \hline
   \multirow{2}{*}{ $U$ [m/s]} & \multicolumn{3}{|c|}{$\delta$ [mm]}  \\ \cline{2-4}
      & Exp. & XFOIL &  $\frac{\vert\mathrm{Exp.} - \mathrm{XFOIL\vert}}{\mathrm{Exp.}}$  \\ \hline
    10  &\{10.2, 12.0\} &\{11.3, 15.4\} &\{ 10.8\%, 28.3\% \}  \\
     30 &\{8.8, 9.5\}&\{10.5, 12.5\} &\{19.3\%, 31.6\%  \} \\
    \hline 
    \end{tabular}
    \caption{Comparison between XFOIL and experimental measurements of  $\delta$ on the suction side for different angles of attack for the NACA~63018 airfoil. The values between curly brackets correspond to the boundary layer parameters for $\alpha_\mathrm{eff}=$~\{0\textdegree, 3\textdegree\}.}
    \label{tab:delta_NACA63018}
\end{table}

\section{Wind turbine blade geometry}\label{app:wt_blade}
The distribution of the geometry parameters of the SWT-2.3-93 wind turbine blade, i.e. airfoil and chord, and flow conditions, i.e., apparent velocity ($U_\mathrm{app}$), and angle of attack ($\alpha$) needed to predict WPS and later far-field noise are shown in Table~\ref{tab:WT_geometry}.
\begin{table}[h!]
\centering
    \begin{tabular}{|c|c|c|c|c|}
    \hline
    Airfoil & Chord [m]& $U_\mathrm{app}$ [m/s] & $Re/10^6$ [-]& $\alpha$ [\textdegree] \\ \hline
    FFA-W3-301 & 3.4 & 26.7&6.1 & 0.28 \\
    FFA-W3-241 & 2.7& 42.47&7.5 &  -1.10\\
    NACA 63-221 & 2.0& 55.0& 7.5&  0.22\\
    NACA 63-218 & 1.6& 64.6& 6.7&  1.12\\
    NACA 63-218 & 1.2& 71.9& 5.7&  1.70\\
    NACA 63-218 & 0.9& 77.6& 4.7&  2.10\\
    NACA 63-218 & 0.7& 81.7& 3.6&  2.20\\
    \hline 
    \end{tabular}
    \caption{Distribution of the conditions for the WPS prediction of the SWT-2.3-93 wind turbine.}
    \label{tab:WT_geometry}
\end{table}

\bibliographystyle{elsarticle-num-names} 
\bibliography{cas-refs}



 
\end{document}